\documentclass[10pt,twocolumn,letterpaper]{article}

\usepackage{iccv}              

\usepackage{graphicx}
\usepackage{lipsum}
\usepackage{multirow}
\usepackage{bm}
\usepackage{amsmath}
\usepackage{mathrsfs}
\usepackage{pifont}
\usepackage{color}
\usepackage{tabularx}
\usepackage{array} 
\usepackage[linesnumbered,ruled,lined]{algorithm2e}
\SetKwRepeat{Do}{Do}{Until}

\newcommand{\figref}[1]{Fig.~\ref{#1}}
\newcommand{\reqref}[1]{Eq.~\eqref{#1}}
\newcommand{\secref}[1]{Sec.~\ref{#1}}
\newcommand{\tabref}[1]{Table~\ref{#1}}
\definecolor{americanrose}{rgb}{1.0, 0.01, 0.24}

\usepackage{xspace}
\makeatletter
\DeclareRobustCommand\onedot{\futurelet\@let@token\@onedot}
\def\@onedot{\ifx\@let@token.\else.\null\fi\xspace}
\def\eg{\emph{e.g}\onedot} 
\def\ie{\emph{i.e}\onedot} 
 
\def\etc{\emph{etc}\onedot} 
 
\def\etal{\emph{et al}\onedot}
\makeatother

\definecolor{iccvblue}{rgb}{0.21,0.49,0.74}
\usepackage[pagebackref,breaklinks,colorlinks,allcolors=iccvblue]{hyperref}

\def\paperID{2799} 
\def\confName{ICCV}
\def\confYear{2025}

\title{Asynchronous Event Error-Minimizing Noise for Safeguarding Event Dataset}

\author{
Ruofei Wang$^1$\quad 
Peiqi Duan$^{2,3}$\quad 
Boxin Shi$^{2,3}$\quad 
Renjie Wan$^{1\ast}$\\
$^1$Department of Computer Science, Hong Kong Baptist University \\ 
\resizebox{\textwidth}{!}{$^2$State Key Laboratory of Multimedia Information Processing, School of Computer Science, Peking University} \\
\resizebox{\textwidth}{!}{$^3$National Engineering Research Center of Visual Technology, School of Computer Science, Peking University} \\
ruofei@life.hkbu.edu.hk, \{duanqi0001, shiboxin\}@pku.edu.cn, renjiewan@hkbu.edu.hk
}

\begin{document}
\twocolumn[{%
\maketitle
\renewcommand\twocolumn[1][]{#1}%

\begin{center}
    \captionsetup{type=figure}
    \includegraphics[width=\linewidth]{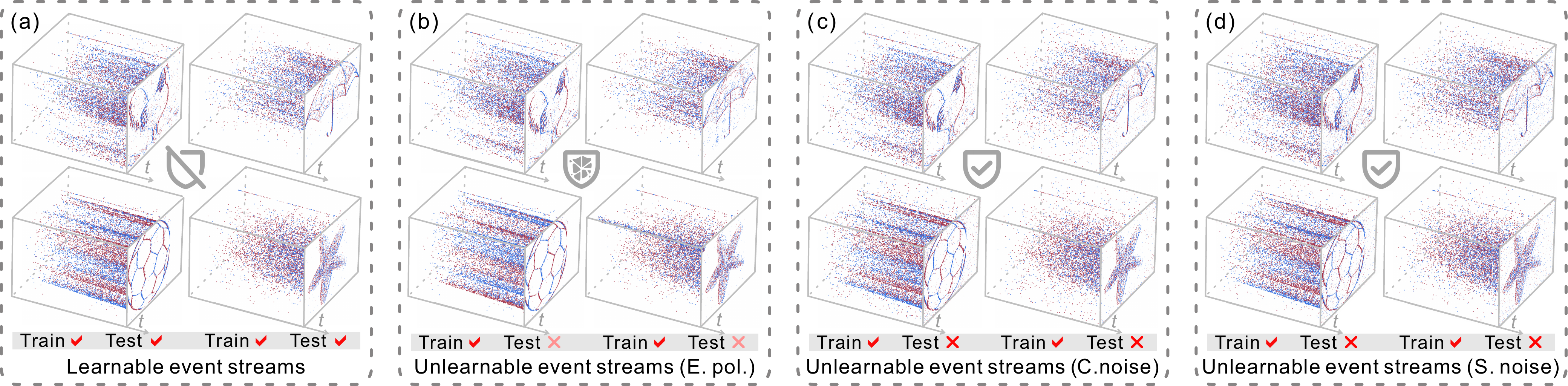}
   \caption{Comparison between the learnable event streams~(\ie, clean data), unlearnable event streams generated by event pollution~(E. pol.), and our $\text{UE}\textsc{v}\text{s}$, including class-wise noise~(C. noise) and sample-wise noise~(S. noise). 
   In case (b), we degrade the quality of these event streams by polluting the coordinates, timestamps, polarity, and adding new random events (reading order).
   } 
\label{fig:motivation}
\end{center}%
}]
{\let\thefootnote \relax \footnote{$^\ast$Corresponding author. This work was carried out at Renjie Group.}}
\begin{abstract}

With more event datasets being released online, safeguarding the event dataset against unauthorized usage has become a serious concern for data owners. 
Unlearnable Examples are proposed to prevent the unauthorized exploitation of image datasets. However, it's unclear how to create unlearnable asynchronous event streams to prevent event misuse.
In this work, we propose the first unlearnable event stream generation method to prevent unauthorized training from event datasets. 
A new form of asynchronous event error-minimizing noise is proposed to perturb event streams, tricking the unauthorized model into learning embedded noise instead of realistic features. To be compatible with the sparse event, a projection strategy is presented to sparsify the noise to render our unlearnable event streams~(UEvs). 
Extensive experiments demonstrate that our method effectively protects event data from unauthorized exploitation, while preserving their utility for legitimate use.
We hope our UEvs contribute to the advancement of secure and trustworthy event dataset sharing. Code is available at: \text{https://github.com/rfww/uevs}.

\end{abstract}    

\section{Introduction}
Recently, the easy availability of various event datasets~\cite{sironi2018hats,gehrig2021dsec,vasudevan2020introduction,kim2021n} has significantly accelerated the development of event vision tasks across various domains, including autonomous driving~\cite{maqueda2018event}, pose estimation~\cite{rudnev2021eventhands,jiang2024evhandpose}, sign-language recognition~\cite{zhang2025evsign}, \etc.
However, a concerning fact is that some datasets are collected for restricted applications, without permission for unauthorized usage~\cite{amir2017low,sironi2018hats}.
The lack of protection mechanisms for event datasets leaves them susceptible to misuse for unauthorized purposes.

To prevent unauthorized dataset usage, a novel concept known as
Unlearnable Examples (UEs)~\cite{huang2021unlearnable} 
is proposed.
Different from data pollution that degrades the dataset quality to avoid misuse~\cite{zhang2024ev,zhang2024event}, UEs incorporates an imperceptible adversarial noise in clean images to make them unlearnable for machine learning models.
This technique perturbs the data in a way that disrupts the learning process of unauthorized models, effectively rendering the data unusable for unauthorized purposes. 
However, for authorized users, these data still retain the original utility and appearance.
Therefore, this approach not only reduces the risk of data misuse but also preserves the dataset's fidelity and integrity, posing an effective way for safeguarding datasets.

Though UEs achieve great success in protecting image datasets~\cite{fu2022robust,ren2023transferable,ye2024ungeneralizable}, how to generate effective unlearnable asynchronous event dataset is unclear.
As illustrated in \figref{fig:motivation}~(a), event data comprises a sequence of asynchronous events (\ie, $(x,y,t,p)$), where $(x,y)$ represents the $x$-$y$ coordinates, $t$ denotes the timestamp, and $p$ indicates the polarity. This special characteristic creates an event stream that resembles spatio-temporal point clouds rather than conventional 2D images~\cite{duan2021eventzoom}.
Therefore, it is impossible to employ existing image UEs methods~\cite{huang2021unlearnable,jiang2023unlearnable,ye2024ungeneralizable} to safeguard asynchronous event datasets. 

Owing to the unique characteristics of event data, there are two main challenges in achieving the unlearnable functionality of event data.
First, event data are distinguished by the binary polarity: $p = \pm1$. It creates a highly discrete pattern distinct from the image samples used in previous UEs. 
Directly incorporating noise based on previous settings can diminish the effectiveness, as the noise contains values outside the binary pattern. This raises a significant challenge: generating effective error-minimizing noise that aligns with the discrete nature of event data. 
Second, current event vision tasks need to transfer the event data into the event stack and then feed it into the downstream DNN models. However, these stacks prevent noise from being injected directly into the event stream. Thus, effectively injecting noise into the event stream presents another significant challenge.

To this end, we propose the \underline{U}nlearnable \underline{Ev}ent \underline{s}tream~($\text{$\text{UE}\textsc{v}\text{s}$}$) method with the Event Error-Minimizing Noise~($\text{E}^2\text{MN}$) to safeguard event datasets.
\textbf{First}, we formulate an event-based optimization function to generate $\text{E}^2\text{MN}$.
Specifically, our $\text{E}^2\text{MN}$ includes two forms: class-wise noise~(\figref{fig:motivation}~(c)) and sample-wise noise~(\figref{fig:motivation}~(d)), with the former generated on the class-by-class basis and the latter on case-by-case basis.
These two forms all aim to build a shortcut between the input sample and target labels to prevent the model from learning real semantic features.
\textbf{Second}, we propose an adaptive projection mechanism to sparsify the generated noise, resulting in the projected noise that can be compatible with event stacks seamlessly.
Thereby, the possibility of our $\text{E}^2\text{MN}$ being detected by malicious users is reduced.
\textbf{Third}, we propose a new retrieval strategy to reconstruct the event stream from its corresponding representation. 
This strategy ensures that $\text{E}^2\text{MN}$ is effectively transferred from unlearnable event stacks to unlearnable event streams.

Our main contributions are as follows: 
\begin{itemize}
    \item We present the first Unlearnable Event stream~($\text{UE}\textsc{v}\text{s}$) with an event error-minimizing noise~($\text{E}^2\text{MN}$) to prevent unauthorized usage of the valuable event dataset. 
    
    \item We propose an adaptive projection mechanism to sparsify the noise $\text{E}^2\text{MN}$, enabling the noise to be compatible with the original event stack.
    Subsequently, a retrieval strategy is introduced to reconstruct the unlearnable event stream from its corresponding event stack.  
    
    \item We conduct extensive experiments to demonstrate that our $\text{E}^2\text{MN}$ outperforms various event pollution operations~(\figref{fig:motivation} (b)) in terms of effectiveness, imperceptibility, and robustness.
\end{itemize}

\section{Related work}
\subsection{Event vision}
Event vision~\cite{duan2021guided,ma2023deformable,duan2023eventaid} has gained increasing popularity recently due to the advantages brought by the event data~\cite{jiang2024complementing,liu2024seeing}, which consists of a series of asynchronous events by recording the change of pixel-wise brightness~\cite{gallego2020event}. Compared with conventional images, event data shows merits in terms of high dynamic range and temporal resolution, low time latency and power consumption~\cite{sironi2018hats,gehrig2019end}.
The easily available and diverse event datasets~\cite{orchard2015converting,li2017cifar10,amir2017low,gehrig2021dsec,kim2021n} have been crucial for the increasing attention to various event vision tasks.
Researchers can effortlessly ensure their event data is compatible with current deep vision models via utilizing existing event representation methods~\cite{lagorce2016hots,gehrig2019end,huang2023eventpoint}, thereby effectively facilitating various explorations in event-based learning.
Millerdurai \etal~\cite{millerdurai2024eventego3d} propose employing an egocentric monocular event camera for 3D human motion capture, which addresses the failure of RGB cameras under low lighting and fast motions.
Sun \etal \cite{sun2022ess} propose an unsupervised domain adaption method for semantic segmentation on event data, which motivates the segmentor to learn semantic information from labeled images to unlabeled events.
Moreover, event-based studies achieve satisfactory performances in pose estimation~\cite{jiang2024evhandpose}, segmentation~\cite{chen2024segment}, deblurring~\cite{cannici2024mitigating,kim2024frequency}, denoising~\cite{duan2021eventzoom}, optical flow~\cite{liu2023tma}, object recognition~\cite{zubic2023chaos}, object tracking~\cite{wang2024event2}, monitoring~\cite{hamann2024low}, \etc. 

Despite the significant interest and promising performance of event-based learning in various domains, there have been no investigations into preventing the unauthorized exploitation of event datasets. 
The absence of this exploration could result in the valuable event datasets being stolen for illegal purposes, leading to serious threat.

\subsection{Unlearnable examples} Unlearnable Examples~(UEs) are proposed to prevent the unauthorized training of Deep Neural Networks~(DNNs) on some private or protected datasets~\cite{huang2021unlearnable,shan2020fawkes}. 
Generally, UEs are generated through a min-min bilevel optimization pipeline with a surrogate model~\cite{huang2021unlearnable}.
The generated noise is called Error-Minimizing Noise~(EMN) because it gradually reduces losses from the training data. This noise aims to deceive the target model into believing that correct predictions can be made merely based on perturbations, resulting in overlooking semantic features~\cite{huang2021unlearnable,ren2023transferable}.
To improve the robustness of vanilla UEs, Fu \etal ~\cite{fu2022robust} introduce adversarial noise into the EMN against adversarial training. 
He \etal \cite{he2023indiscriminate} propose to generate unlearnable examples based on unsupervised contrastive learning, which extends the supervised unlearnable noise generation into unsupervised learning.
Ren \etal \cite{ren2023transferable} propose Classwise Separability Discriminant, which aims to better transfer the unlearnable effects to other training settings and datasets by enhancing the linear separability.
Meng~\etal \cite{meng2024semantic} propose a deep hiding strategy that adaptively hides semantic images into the latent domain of poisoned samples to generate UEs. 
Although current UEs have shown promising results in image area~\cite{sun2024unseg,liu2024multimodal,liu2024stable}, this technique cannot be directly applied to protect asynchronous event streams due to the different data characteristics.
To this end, our work focuses on exploring the unlearnable event stream to safeguard the event dataset against unauthorized usage.

\begin{figure*}[ht]
    \centering
    \includegraphics[width=\linewidth]{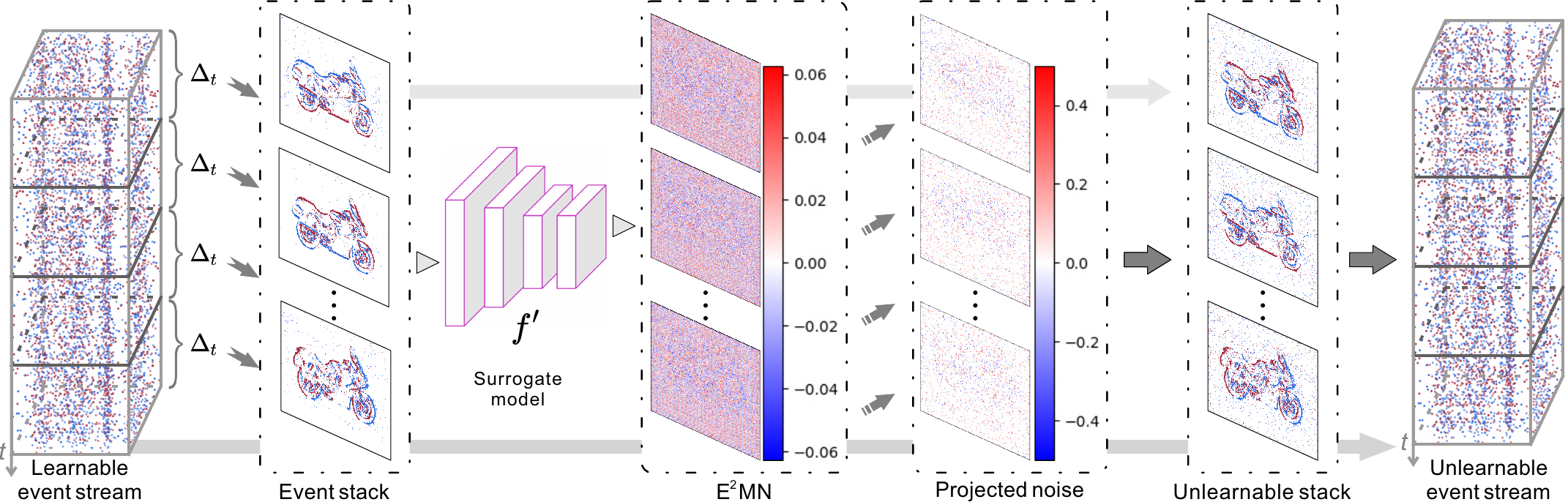}
    \caption{Framework of our $\text{UE}\textsc{v}\text{s}$. 
    $\text{UE}\textsc{v}\text{s}$ converts an event stream to the $C$-channel event stack~($t=C\times\Delta_t$) and then employs a surrogate model $f'$ to generate the event error-minimizing noise~($\text{E}^2\text{MN}$). Subsequently, the $\text{E}^2\text{MN}$ is projected into $\{-0.5, 0, 0.5\}$ and integrated with the target event stack to generate the unlearnable event stack. 
    The final unlearnable event stream can be reconstructed from the unlearnable event stack and original event stream via the proposed retrieval strategy. 
    }
    \label{fig:framework}
\end{figure*}

\section{Methodology}
\subsection{Preliminary of event data}
\label{sec:preliminary}

Given an event data-based classification dataset ${\mathcal{D}}=\{(\mathcal{E}_i,l_i)\}_{i=1}^{N}$, where $\mathcal{E}$, $l$, and $N$ indicate the event stream, the corresponding category label, and the dataset length, respectively. The event stream $\mathcal{E}$ consists of a variety of individual events, recorded as:
\begin{equation}
    \mathcal{E}=\{\textbf{\textit{e}}_k\}_{k=1}^{K}=\{(x_k,y_k,t_k,p_k)\}_{k=1}^{K},
    \label{eq:event}
\end{equation}
where $(x_k,y_k,t_k,p_k)$ indicates the $x$ and $y$ direction coordinates, time stamp, and polarity of the $k$-th event, respectively. $K$ is the length of the event stream $\mathcal{E}$~\cite{gallego2020event}. 
Specifically, an event, $\textbf{\textit{e}}_k$, has occurred when the variation of the log brightness at each pixel exceeds the threshold $\sigma$, \ie, $|\log(x_k,y_k,t_k)-\log(x_{k},y_{k},t_{k-1})|>\sigma$. The polarity $p_k=+1.0$ when the difference between bi-temporal brightness is higher than $+\sigma$. Otherwise, $p_k$ is set to $-1.0$.

\subsection{Our scenario}
We propose $\text{UE}\textsc{v}\text{s}$ to prevent unauthorized training on the valuable event datasets that effectively protect the data owner's interests.
Generally, two parties are mentioned in unlearnable event streams, \ie, the protector and the hacker. 
The protector has full accessibility to the original dataset and generates $\text{UE}\textsc{v}\text{s}$ for his/her private data before the dataset is released. To ensure the unlearnability, the protector can use various surrogate models to generate the event error-minimizing noise. However, he/she cannot touch any information about the models or training skills that the hacker would use. 
As a hacker, he/she has no knowledge about the original datasets and surrogate models. He/she aims to steal these data to train their illegal models with any training tricks.
In this paper, we are the protector to make the collected event dataset unlearnable with the imperceptible noise $\text{E}^2\text{MN}$.

\subsection{Unlearnable event stream}

\paragraph{Overview.} The overall pipeline of our $\text{UE}\textsc{v}\text{s}$ is shown in \figref{fig:framework}. The framework takes the original event stream as input and outputs an unlearnable one.
Based on the fact that an example with higher training loss contains more useful knowledge to be learned~\cite{fu2022robust}, we formulate the pipeline of our Unlearnable Event streams~($\text{UE}\textsc{v}\text{s}$) as:
\begin{equation}
    \arg \min_{\theta} \mathbb{E}_{(\mathcal{E}, l)\in\mathcal{D}}[\min_{\delta}\mathcal{L}(f_\theta'(\mathcal{R}(\mathcal{E})+\bm{\delta}), l)]\ \text{s.t.}\ ||\delta||_{\infty}\leq\epsilon,
    \label{eq:emn}
\end{equation}
where $f'$ denotes a surrogate model used for noise $\delta$ generation, $\mathcal{L}$ indicates the loss function, and $\mathcal{R}$ means event representation. 
\reqref{eq:emn} is a min-min bi-level optimization function, where both constrained optimization items all aim to minimize the loss of the surrogate model. 
Specifically, the inner minimization is employed to find the $L_\infty$-norm bounded noise $\delta$ to improve the model's performance, while the outer optimization finds the optimal parameter $\theta$ to minimize the model’s classification loss. 
Obviously, the core component of $\text{UE}\textsc{v}\text{s}$ is to find the effective noise $\delta$ to minimize the loss of $f'$, thereby suppressing the model from learning real semantic features from input events.

To generate unlearnable event streams, first, we convert the asynchronous event data into a regular event stack; second, we feed this stack into a surrogate model to compute the noise $\delta$; third, we project $\delta$ to be compatible with the binary-polarity event stack; and fourth, we reconstruct the unlearnable event stream from the unlearnable event stack.

\paragraph{Event representation.}
To be compatible with existing vision models~\cite{he2016deep,simonyan2015very,tan2019efficientnet}, the event stream needs to be transferred into image-like stacks and then input as images for the following predictions~\cite{rebecq2019high,berl2023lorenzo}. 
To meet the regular input requirement of the surrogate model, we propose binning events into a $C$-channel event stack. We set $C$ with a large value to prevent possible event corruption since the single or three-channel maps may easily lead to the coverage of previous events. Each channel of this stack consists of polarities of those events distributed in the divided time bin~$\Delta_t$. Hence, our event stack only contains three types of values at each coordinate: $\{0, 0.5, 1.0\}$, denoting an event with $\text{polarity}=-1$, no event, and an event with $\text{polarity}=+1$, respectively. Note that the event data cannot be reconstructed from the event stack if any values other than these two specific values are used.

\paragraph{Noise generation.}
The event stack is then fed into the surrogate model $f'$ used for the noise generation. 
The training steps of $f'$ are limited since training the surrogate model with larger steps would lead to ineffective event error-minimizing noise optimization. 
Thus, we train the parameter $\theta$ of $f'$ with only $M$ iterations first and then optimize $\delta$ across the entire $\mathcal{D}_c$.
This entire bi-level optimization process~(\reqref{eq:emn}) is terminated once the classification accuracy is higher than $\gamma$. 
Essentially, the generated event error-minimizing noise is a kind of easy-to-learn feature~(\ie, the shortcut between input samples and target labels) that guides the model trained on $\text{E}^2\text{MN}$ to ignore their realistic semantics. 
Our event error-minimizing noise consists of two types:
\begin{itemize}
    \item \textbf{Sample-wise noise.} Sample-wise noise $\Delta_s$ is obtained by 
    the first-order optimization method PGD \cite{madry2018towards} as follows:
    \begin{equation}
    \label{eq:adversarial-noise}
    \bm{\mathcal{\hat{E}}}'_{s+1} = \Pi_{\epsilon} \big( \bm{\mathcal{\hat{E}}}'_{s} - \alpha \cdot \text{sign}(\nabla_{\bm{\mathcal{\hat{E}}}} \mathcal{L}(f^\prime(\bm{\mathcal{\hat{E}}}'_{s}), l)) \big),
    \end{equation}
    where, $\bm{\mathcal{\hat{E}}}=\mathcal{R}(\mathcal{\bm{E}})$ indicates the event stack, $s$ denotes the current perturbation step, $\nabla_{\bm{\mathcal{\hat{E}}}} \mathcal{L}(f(\bm{\mathcal{\hat{E}}}'_{s}), l)$ is the gradient of the loss with respect to the $\bm{\mathcal{\hat{E}}}$, $\Pi_\epsilon$ is a projection function that clips the perturbed event stack back to the $\epsilon$-ball around the event stack $\bm{\mathcal{\hat{E}}}$ when it goes beyond, and $\alpha$ is the step size. $\mathcal{L}$ is the cross entropy loss function.
    The generated event error-minimizing noise is achieved by: $\bm\delta_i = \bm{\mathcal{\hat{E}}}_i' - \bm{\mathcal{\hat{E}}}_i$ and calculates it case-by-case to get $\Delta_s=\{\bm{\delta}_1, \bm{\delta}_2,\cdots,\bm{\delta}_N\}$.
    The output $\bm{\mathcal{\hat{E}}}'$ is not our final unlearnable event stack since it does not exhibit a binary-polarity characteristic, which is crucial for the reconstruction of unlearnable events.
    We need to project this noise into a sparse pattern that is compatible with the original event stack to generate our unlearnable event streams.

    \item \textbf{Class-wise noise.} 
    Since sample-wise noise depends on each input event, memory consumption would increase sharply as the dataset scale grows. Therefore, we introduce the class-wise noise: $\Delta_c=\{\bm{\delta}_{l_1},\bm{\delta}_{l_2}\cdots,\bm{\delta}_{l_N}\}$, which applies the same noise to different events within the same class.
    For $\Delta_c$, we firstly initialize the noise in the class-level and then employ \reqref{eq:adversarial-noise} to optimize the noise $\bm\delta_{l}$ with all event streams sampled from the class $l$.
    Class-wise noise shows advantages over sample-wise noise in terms of practicality and efficiency. 
    It can be easily applied to newly captured event streams and has a smaller noise scale.  Same to the sample-wise noise, class-wise noise cannot be directly used to generate unlearnable event streams, which also needs to be projected into a sparse form for $\text{UE}\textsc{v}\text{s}$ generation.

\end{itemize}

\paragraph{Noise projection.} As demonstrated in \secref{sec:preliminary}, the event stream consists solely of polarities of $-1$ and $+1$, which are normalized to $0$ and $1$ in event stack, receptively. Therefore, we need to project the generated event error-minimizing noise into a sparse and event-friendly pattern to facilitate the subsequent generation of unlearnable event streams. Here, we propose projecting the generated noise into $\{-0.5, 0.0, +0.5\}$ as follows:

\begin{equation}
\label{eq:projection}
 \mathbf{P}(\delta)=\left\{
\begin{aligned}
-0.5, &\quad \text{if}\ \ \delta_{i, j}<\mu-\tau\times\pi, \\
 0.0, &\quad \text{if}\ \ \mu-\tau\times\pi\leq\delta_{i, j}\leq\mu+\tau\times\pi, \\
+0.5, &\quad \text{if}\ \ \delta_{i, j}>\mu+\tau\times\pi,
\end{aligned}
\right.
\end{equation}
where, the $\mu$ and $\pi$ denote the mean and $1/2$ bound of noise $\delta$, $\tau$ is a balancing parameter. 

$\tau$ imposes great influence on two criteria of our $\text{UE}\textsc{v}\text{s}$, \ie, effectiveness and imperceptibility. Effectiveness means that the projected noise still owns the unlearnable functionality for event streams. Imperceptibility denotes that the injected noise is imperceptible to users. A higher $\tau$ benefits better imperceptibility but corrupts the effectiveness. 
To ensure the effectiveness of the projected noise, we enforce the surrogate model to discriminate between the unlearnable features and clean features during model training. Thereby, the model can generate a strong shortcut during noise optimization, preventing the model from learning real semantic features. We reformulate the loss function of the outer optimization in \reqref{eq:emn} as:
\begin{equation}
\label{eq:loss_surrogate}
    \mathcal{L}^*=\lambda_1\mathcal{L}+\lambda_2\mathcal{L}_{s}. 
\end{equation}
$\mathcal{L}_{s}$ is the similarity loss that aims to enlarge the difference between clean and unlearnable features. 

\noindent\textbf{Event reconstruction.} After getting the projected event error-minimizing noise, we incorporate this noise into $\mathcal{\hat{E}}$ to generate the unlearnable counterpart. Note that what we obtain here is only the unlearnable event stack.
We need to reconstruct the unlearnable event stream from this stack. 
Therefore, a retrieval strategy is proposed to recover the compressed dense temporal information when converting event data into event stacks.
%
For an event within the unlearnable event stack $\mathcal{\hat{E}}$,
\textbf{1)} if it is recorded in original event stream $\mathcal{E}$ then search for the time stamp $t$ in $\mathcal{E}$ to assign it directly;
\textbf{2)} if it is a newly generated event, initialize a new adaptive time stamp based on the selected $\Delta_t$ to set it.
Based on this strategy, the generated unlearnable events can be transferred from unlearnable event stacks into event streams, as shown in \figref{fig:motivation}. 
The overall pipeline of our $\text{UE}\textsc{v}\text{s}$ is illustrated in Algorithm \ref{algorithm}. $\mathcal{R}(\cdot)$ and $\mathcal{R}'(\cdot)$ denote event representation and event reconstruction, respectively.

\subsection{Implementation details}
Following \cite{huang2021unlearnable}, we employ the ResNet18~\cite{he2016deep} as the surrogate model. The SGD optimizer is employed with a learning rate of $10^{-4}$ and a momentum of 0.9. An exponential learning rate scheduler with a gamma of 0.9 is used. The iteration number $M$ and epoch number are set to 10 and 30, respectively. The batch size is 16. The $C$ in the event stack is set to $16$. In noise generation, the step number $S$, parameter $\epsilon$, and step size $\alpha$ are 10, 0.5, and 0.8/255, respectively. The balancing ratio $\tau$ is $3/4$. The accuracy $\gamma$ is 0.99. We train all models on a single NVIDIA V100 GPU.

\begin{algorithm}[t]
\caption{$\text{UE}\textsc{v}\text{s}$ Perturbation Algorithm.}\label{algorithm}
\KwIn{Surrogate model $f'_\theta(\cdot)$, clean dataset $\mathcal{D}_c$, optimization number $M$, accuracy $\gamma$.}
\KwOut{ Perturbation $\bm{\delta}$, Unlearnable dataset $\mathcal{D}_u$.}
$\bm{\delta}\leftarrow \textit{randomization}$\;
\Do{$acc\ \text{higher than}\ \gamma$}
{
\For{$i \text{ in } \text{range}(M)$}{$(\mathcal{E}, l)=Next(\mathcal{D}_c)$\;
$\theta \leftarrow \theta-\nabla_\theta(\mathcal{L}^*(f'_\theta(\mathcal{R}(\mathcal{E})+[\bm{\delta}_{i}\ /\ \bm{\delta}_{l_i}]), l))$
\quad\text{$\triangleright$  \color{gray} Sample\ or\ class -wise noise:}$\ \bm{\delta}_i/\bm{\delta}_{l_i}$
}

\For{$(\mathcal{E}, l)\ \text{in}\ \mathcal{D}_c$}{$\bm{\delta}=\mathbb{P}(\mathcal{E}, l, f', \delta, \theta)$;
\quad\text{$\triangleright$ \color{gray} $\mathbb{P}$erturbed by \reqref{eq:adversarial-noise}}

$Clip(\bm{\delta}, -\epsilon, +\epsilon)$\;}
$acc = \textit{evaluation}(\mathcal{D}_c, f', \delta, \theta)$\;}

\For{$(\mathcal{E}_j, l_j)\ in\ \mathcal{D}_c$}{
$\delta'=\mathbf{P}(\delta\leftarrow[\bm{\delta}_j\ or\ \bm{\delta}_{l_j}])$;\quad\text{$\triangleright$ \color{gray} \reqref{eq:projection}}

$\mathcal{E'}=\mathcal{R'}\big(\text{Clip}(\mathcal{R}(\mathcal{E}_j)+\delta',\ 0,\ 1)\big)$\;

$\mathcal{D}_u.append(\bm{(}\mathcal{E}_j', l_j\bm{)})$;\quad\text{$\triangleright$ \color{gray} Unlearnable event}}

\end{algorithm}

\section{Experiment}
\subsection{Setup}
\paragraph{Dataset.} Four event-based datasets are used in our experiments, including N-Caltech101~\cite{orchard2015converting}, CIFAR10-DVS~\cite{li2017cifar10}, DVS128 Gesture~\cite{amir2017low}, and N-ImageNet~\cite{kim2021n}. 
Due to the limited computing resources, the first 10 classes are sampled from N-ImageNet~\cite{kim2021n} for experiments.
Detailed information is shown in \tabref{tab:dataset} of the Supplementary Material.

\paragraph{Baseline.} To evaluate the effectiveness, we implement several event pollution operations~(see \figref{fig:motivation}~(b)) as our baselines, including coordinate shifting~(CS), time stamp shifting~(TS), polarity inversion~(PI), manual pattern injection~(MP), and area shuffling~(AS). To test the generalizibility of $\text{UE}\textsc{v}\text{s}$, we introduce ResNet18~(RN18)~\cite{he2016deep}, ResNet50~(RN50)~\cite{he2016deep}, VGG16~(VG16)~\cite{simonyan2015very}, DenseNet121~(DN121)~\cite{huang2017densely}, EfficientNet-B1~(EN-B1)~\cite{tan2019efficientnet}, $\text{ViT}_\text{B}$~\cite{dosovitskiy2020image}, and $\text{Swin}_\text{B}$~\cite{liu2021swin} as baselines in our experiment.

\begin{figure*}[ht]
    \centering
    \includegraphics[width=\linewidth]{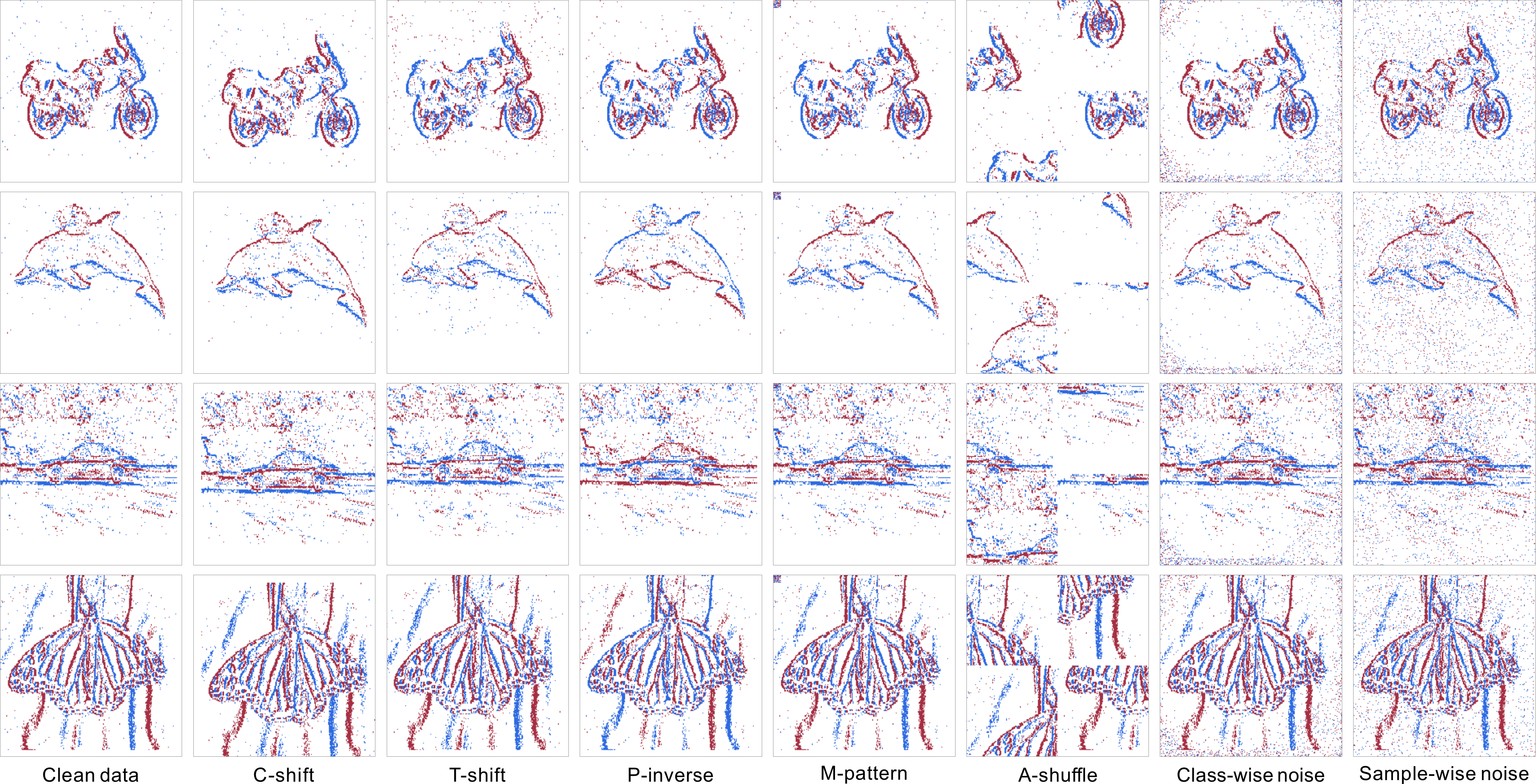}
    \caption{Visualization results of various noise forms on the N-Caltech101 dataset~\cite{orchard2015converting}. Blue/Red points denote the events with $p=+1$/$-1$. 
    Our noise $\text{E}^2\text{MN}$ does not corrupt the target objects, and the noise distributed in the background region appears quite realistic.
    }
    \label{fig:result}
\end{figure*}

\paragraph{Metric.} Following the existing UEs method~\cite{huang2021unlearnable}, we use the test accuracy to evaluate the data protection ability of the proposed method. The lower the test accuracy, the stronger the protection ability. To assess the imperceptibility of the generated noise, we employ PSNR, SSIM, and MSE to measure the visual perception on the event stack~\cite{duan2021eventzoom}.

\subsection{Evaluation}
\paragraph{Visualization comparison} of aforementioned various noise forms is shown in \figref{fig:result}. We implement five kinds of event distortion operations as baselines to evaluate our $\text{UE}\textsc{v}\text{s}$. The coordinate shifting~(CS) only introduces an offset in $x$ and $y$ directions without adding any additional noise, as shown in the 2nd column. We put off the time stamps of each event by a fixed time bin to achieve the TS, which shows that some new events are introduced in the current representations. Reverse the polarity of all events~(PI) is another noise form that shows a noticeable difference between the clean data and unlearnable counterpart in the 4th column. MP is implemented by injecting special event streams manually, which shows an obvious pattern in the top-left corner of the representation. We also shuffle the event in block-level~(AS) to confuse the model training, which shows a noticeable difference between the clean data and unlearnable one poisoned by AS. 
Our class-wise noise and sample-wise noise are shown in the last two columns of \figref{fig:result}.
Detailed imperceptibility evaluation of each kind of noise is shown in \tabref{tab:visual}. Injecting a small manual pattern~(MP) can achieve a good imperceptibility. But our $\Delta_c$ and $\Delta_s$ achieve the best overall performance in imperceptibility and unlearnability. The PSNR of $\Delta_c$ and $\Delta_s$ are 20.39 and 18.22, respectively.

\paragraph{Quantitative results} of various unlearnable noise forms on the four datasets are shown in \tabref{tab:experiment}. We can find that our sample-wise and class-wise noise achieve a noticeable accuracy drop among the whole classifiers compared with the baselines. Shifting the coordinates or time stamps of the event cannot achieve a reliable unlearnable effect. For example, on the DVS128 Gesture dataset, ResNet18 achieves an improvement by 1.21\% when train the classifier on $\mathcal{D}_u$ poisoned by CS than the clean dataset $\mathcal{D}_c$. 
Setting the polarity of each event to inverse can confuse the model to learn semantic features, where PI achieves the accuracy drop by $40.87\%$ on the N-Caltech101 dataset. However, this noise can be removed easily via simple data augmentation skills. 
Injecting a manual pattern~(MP) into the event stream to build a shortcut between the event data and target label is inspired by backdoor attacks~\cite{wang2025event}. We randomize patterns for each class to poison the event data that influences the accuracy of ResNet18 from $65.19\%$ to $36.12\%$ on the CIFAR10-DVS dataset. On N-ImageNet, the MP achieves an obvious accuracy drop by 45.40 since it is easier to be captured by deep models than the real complex features of input data.
Shuffling the event in block-level~(AS)~(see the 6th column of \figref{fig:result}) cannot protect the event stream effectively, which achieves the accuracy by $48.25\%$, $39.81\%$, $58.68\%$, and $28.00\%$ on N-Caltech101, CIFAR10-DVS, DVS128 Gesture, and N-ImageNet, respectively. 

\begin{table}[]
    \centering
        \caption{Imperceptibility evaluation of various noise forms.}
    \resizebox{\linewidth}{!}{
    \begin{tabular}{cccccccc}
    \toprule
    &CS&TS&PI&MP&AS&$\Delta_c$ & $\Delta_s$ \\
    \midrule
        PSNR & 12.19& 13.07& 16.76& 28.82& 11.75& 20.36& 18.22\\
        SSIM &  0.316& 0.418& 0.996& 0.997& 0.245 & 0.791& 0.571\\
        MSE & 0.182 & 0.143 & 0.067 & 0.069 & 0.212 & 0.028 & 0.040 \\
        \bottomrule
    \end{tabular}}

    \label{tab:visual}
\end{table}

\begin{table*}[t]
    \centering
    \footnotesize
    \caption{The testing accuracy (\%) of various DNN classifiers trained on the clean training sets ($\mathcal{D}_c$) and their unlearnable ones ($\mathcal{D}_u$) generated by five kinds of event pollutions, sample-wise noise ($\Delta_s$), and class-wise noise ($\Delta_c$). 
    }
    \begin{tabular}{p{1.2cm}<{\centering}|l|p{1cm}<{\centering}p{1.8cm}<{\centering}|p{1cm}<{\centering}p{1.8cm}<{\centering}|p{1cm}<{\centering}p{1.8cm}<{\centering}|p{1cm}<{\centering}p{1.8cm}<{\centering}}

    \toprule
    \multirow{2}{*}{Noise Form} & \multirow{2}{*}{Model} & \multicolumn{2}{c|}{N-Caltech101} &  \multicolumn{2}{c|}{CIFAR10-DVS}& \multicolumn{2}{c}{DVS128 Gesture}& \multicolumn{2}{c}{N-ImageNet} \\
    
    & & $\mathcal{D}_c$ & $\mathcal{D}_u$ & $\mathcal{D}_c$ & $\mathcal{D}_u$ & $\mathcal{D}_c$ & $\mathcal{D}_u$& $\mathcal{D}_c$ & $\mathcal{D}_u$  \\
    \midrule
    CS & \multirow{5}{*}{Res18} &\multirow{5}{*}{78.32} 
    &$50.43_{({\color{red}-27.79})}$ &\multirow{5}{*}{65.19}
    &$46.80_{({\color{red}-18.39})}$ &\multirow{5}{*}{74.14} 
    &$75.35_{({\color{red}+01.21})}$ &\multirow{5}{*}{56.60}  &$42.60_{({\color{red}-14.00})}$\\
    
    TS & & &$37.54_{({\color{red}-40.87})}$ & &$35.34_{({\color{red}-29.85})}$ & &$72.92_{({\color{red}-01.22})}$ & &$41.20_{({\color{red}-15.40})}$\\
    
    PI & & &$40.78_{({\color{red}-37.45})}$ & &$22.72_{({\color{red}-42.47})}$ & &$31.94_{({\color{red}-42.20})}$ & &$41.80_{({\color{red}-4.80})}$\\

    MP & & &$50.26_{({\color{red}-28.06})}$ & &$36.12_{({\color{red}-29.07})}$ & &$68.06_{({\color{red}-06.08})}$ & &$11.20_{({\color{red}-45.40})}$\\
    
    AS & & &$48.25_{({\color{red}-30.07})}$ & &$39.81_{({\color{red}-25.38})}$ & &$58.68_{({\color{red}-15.46})}$ & &$28.00_{({\color{red}-28.00})}$\\
    \midrule
    \multirow{6}{*}{$\Delta_c$} 
    & RN18 &78.52 &$01.90_{({\color{red}-76.62})}$  &65.22 &$22.33_{({\color{red}-42.89})}$ 
    &74.31 & $14.93_{({\color{red}-59.38})}$ &56.80 &$10.00_{({\color{red}-46.80})}$\\
    
    & RN50 &80.64 &$04.88_{({\color{red}-75.76})}$  &67.67 & $18.83_{({\color{red}-48.84})}$
    &78.47&$24.31_{({\color{red}-54.16})}$ &61.60 & $10.00_{({\color{red}-51.60})}$\\
    
    & VG16 &71.63 &$03.68_{({\color{red}-67.95})}$  &67.77 &$30.19_{({\color{red}-37.58})}$ 
    &81.94&$20.49_{({\color{red}-61.45})}$ &56.20 &$04.00_{({\color{red}-52.20})}$\\
    
    & DN121 &75.59 & $02.18_{({\color{red}-73.31})}$  &62.72  &$24.37_{({\color{red}-38.35})}$ 
    &74.31&$22.22_{({\color{red}-52.09})}$ &59.20 &$08.80_{({\color{red}-50.40})}$\\
    
    & EN-B1 &74.44 &$12.81_{({\color{red}-61.63})}$  &72.72 &$23.50_{({\color{red}-49.22})}$ 
    &62.50&$19.10_{({\color{red}-43.40})}$ &56.20 &$06.20_{({\color{red}-50.00})}$\\
    
    & $\text{ViT}_\text{B}$ &44.69 &$01.55_{({\color{red}-43.14})}$  &45.44 &$09.51_{({\color{red}-35.93})}$
    &66.67&$26.74_{({\color{red}-39.93})}$ &41.00 &$10.00_{({\color{red}-31.00})}$\\
    
    & $\text{Swin}_\text{B}$ &90.70 &$00.52_{({\color{red}-90.18})}$  & 75.24&$29.03 _{({\color{red}-46.21})}$
    &83.29&$28.12_{({\color{red}-55.17})}$ &72.00 &$10.00_{({\color{red}-62.00})}$\\

    \midrule
    \multirow{6}{*}{$\Delta_s$} 
    & RN18 &78.12 & $00.52_{({\color{red}-77.60})}$  
    & 65.15& $15.51_{({\color{red}-49.64})}$ 
    &73.96 & $14.54_{({\color{red}-59.42})}$ &56.40 &$10.20_{({\color{red}-46.20})}$\\
    
    & RN50 &80.30 &$06.09_{({\color{red}-74.21})}$ 
    & 67.86& $12.62_{({\color{red}-55.24})}$ &79.17&$20.03_{({\color{red}-59.14})}$ &61.20 &$16.80_{({\color{red}-44.40})}$\\
    
    & VG16 &72.03 &$14.13_{({\color{red}-57.90})}$
    &  67.18& $27.86_{({\color{red}-39.32})}$ &81.94&$26.39_{({\color{red}-55.55})}$ &55.40 & $14.80_{({\color{red}-40.60})}$\\
    
    & DN121 &75.30 &$11.49_{({\color{red}-63.85})}$
    &  64.27& $15.73_{({\color{red}-48.54})}$
    &74.31&$20.83_{({\color{red}-53.48})}$ &59.20 & $11.40_{({\color{red}-47.80})}$\\
    
    & EN-B1&73.81 &$17.98_{({\color{red}-55.83})}$
    & 72.82& $14.95_{({\color{red}-57.87})}$ 
    & 64.93&$16.32_{({\color{red}-48.61})}$ &55.60 & $11.40_{({\color{red}-44.20})}$\\
    
    & $\text{ViT}_\text{B}$ &44.73 &$01.09_{({\color{red}-43.64})}$
    &  47.38& $26.02_{({\color{red}-21.36})}$ &67.01&$30.21_{({\color{red}-36.80})}$ &40.40 &$08.80_{({\color{red}-31.60})}$\\
    
    & $\text{Swin}_\text{B}$ &90.70 &$21.14_{({\color{red}-69.56})}$ &75.44 & $27.18_{({\color{red}-48.26})}$
    &83.68 & $17.01_{({\color{red}-66.67})}$ &71.20 &$09.00_{({\color{red}-62.20})}$\\
    
    \bottomrule
    
    \end{tabular}
    \label{tab:experiment}
\end{table*}

On the contrary, our sample-wise noise and class-wise noise all achieve credible unlearnable performance on four datasets. Since ResNet18 is selected as the surrogate model in our experiment, the best unlearnable effectiveness is obtained when a malicious user trains ResNet18 on the unlearnable datasets. 
To evaluate the generalizability of the $\text{UE}\textsc{v}\text{s}$, we have trained a series of popular DNN classifiers on the clean data and our unlearnable counterpart. As shown in \tabref{tab:experiment}, the class-wise noise can deceive almost all of the classifiers to learn shortcuts on N-Caltech101, where the accuracy of these classifiers is less than $5\%$ except EfficientNet-B1. Moreover, it also achieves a great accuracy drop higher than $35\%$ on both CIFAR10-DVS and DVS128 Gesture datasets. For sample-wise noise, every data has a different low-error noise, which noise can reduce the loss during the optimization process of the model, which leads to the event data unlearnable. In \tabref{tab:experiment}, sample-wise noise $\Delta_s$ achieves the biggest accuracy drop by $77.60\%$, $57.87\%$, and $66.67\%$ on four datasets, respectively.

\subsection{Ablation study}
\label{sec:ablation}
\begin{figure*}[ht]
    \centering
    \includegraphics[width=\linewidth]{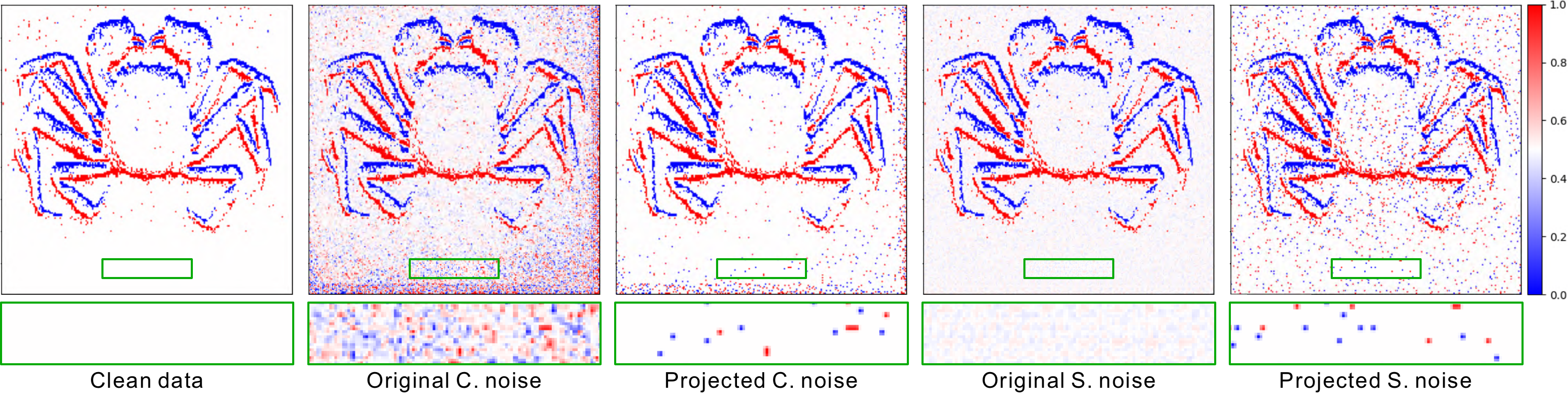}
    \caption{Visualization results of clean data representation, unlearnable representations generated by class-wise noise and sample-wise noise with and without projection, respectively. Details are zoomed in the green box.}
    \label{fig:projection}
\end{figure*}

\paragraph{Similarity loss.} To ensure the event error-minimizing noise can still prevent model learning after being projected, we introduce the similarity loss to enlarge the distance between the clean and unlearnable stacks during the surrogate model training. Detailed results are shown in \tabref{tab:abl_exp}. Compared with $\textbf{E1}$ in \tabref{tab:abl_exp}, the effectiveness of our sample-wise noise~($\textbf{E2}$) is decreased when we remove the similarity loss. Minimizing the similarity between the clean and unlearnable stacks is to improve the difference between the shortcut and real semantic features of input samples under the correct classification, avoiding possible feature learning. As shown in \figref{fig:projection}, this similarity supervision is crucial for ensuring the unlearnability of our $\text{UE}\textsc{v}\text{s}$ due to the big difference between the original noise and projected noise.

\paragraph{Mixed noise.}
In the main experiment, we have conducted exploration about the sample-wise noise and class-wise noise, respectively. According to detailed results, we can find that sample-wise and class-wise noises achieve the best performance on different datasets among various classifiers. Here, we propose a mixed counterpart to evaluate the effectiveness of our $\text{UE}\textsc{v}\text{s}$, which is beneficial for personalized utilization and preventing noise exposure. Since it's hard to optimize these two kinds of noises simultaneously, we mix the obtained two noises as the mixed counterpart instead of training it from scratch. Specifically, we design two modes to generate the mixed noise.
First, we randomly select $\delta$ from $\Delta_c$ and $\Delta_s$ to poison the input sample, denoting as $\Delta_c\vee\Delta_u$. Second, we employ the element-wise addition to generate the unlearnable examples, naming $\Delta_c+\Delta_u$.
\begin{table}[t]
    \centering
        \caption{Results of ablation experiments for $\text{UE}\textsc{v}\text{s}$~(sample-wise noise) on N-Caltech101 dataset.}
    \resizebox{\linewidth}{!}{
    \begin{tabular}{c|c|c|c|c|c|c|c}
    \toprule
         & RN18&RN50&VG16&DN121&EN-B1&$\text{ViT}_B$&$\text{Swin}_B$ \\
         \midrule
        E1 & 0.52&6.09&14.13&11.49&17.98&1.09&21.14\\
        E2 & 3.85&15.22&38.31&10.17&24.18&37.51&27.51\\
        E3 & 5.17&5.40&22.23&6.15&18.78&1.21&16.66\\
        E4 & 5.34&5.17&8.16&5.17&5.86&1.03&0.57\\
        E5 & 13.04& 8.04 &14.99& 10.74 & 16.54 & 9.88 & 25.73 \\
        E6 & 27.74 &33.08 &44.28 &34.92 &27.17 &8.85 &24.41 \\
         \bottomrule
    \end{tabular}}

    \label{tab:abl_exp}
\end{table}

As shown in \tabref{tab:abl_exp}, $\textbf{E3}$ and $\textbf{E4}$ represent the $\Delta_c\vee\Delta_u$ and $\Delta_c+\Delta_u$ respectively, where the proposed event error-minimizing noise remains highly effective. 
Although employing $\Delta_c+\Delta_u$ can achieve a better unlearnable performance, simple addition leads to the low stealthiness of the noise. Randomly selecting noise from $\Delta_c\vee\Delta_u$ provides another possible solution for dataset protection, making our $\text{UE}\textsc{v}\text{s}$ more flexible in practical usage. 

\paragraph{Robustness.} In event-based learning, data augmentation is always adopted to enhance the model's performance. Therefore, studying the robustness of our unlearnable event streams against data augmentation methods is of great importance. Here, we employ the common random shift, crop, flip, and event drop~\cite{gu2021eventdrop} as augmentation techniques in our experiments. As shown in \tabref{tab:abl_exp} \textbf{E5}, these augmentation techniques slightly eliminate the effectiveness of our event error-minimizing noise. This evaluation demonstrates the high practicality of our $\text{UE}\textsc{v}\text{s}$ in various training scenarios.

\paragraph{Different adversarial attacks.}
As shown in \reqref{eq:adversarial-noise}, our $\text{E}^2\text{MN}$ is generated inspired by the PGD attack.
Therefore, exploring the generalization ability of our method encountering other adversarial attack methods~(\eg, FGSM~\cite{goodfellow2014explaining}) becomes natural. For example, we reformulate the \reqref{eq:adversarial-noise} via FGSM as
$\bm{\mathcal{\hat{E}}}' =  \bm{\mathcal{\hat{E}}} - \alpha \cdot \text{sign}(\nabla_{\bm{\mathcal{\hat{E}}}} \mathcal{L}(f^\prime_\theta(\bm{\mathcal{\hat{E}}}), l))$ to induce the model focusing on easy-learned features. Here, we set $\alpha$ to $8/255$ and use the same configurations with \reqref{eq:adversarial-noise} to train the surrogate model. In \tabref{tab:abl_exp}, \textbf{E6} shows the unlearnable performance of our $\text{UE}\textsc{v}\text{s}$ via FGSM on N-Caltech101 dataset. The unlearnable ability of our method is indeed decreased by using FGSM. However, $\text{UE}\textsc{v}\text{s}$ can still prevent unauthorized models from achieving reliable predictions under the protection provided by $\text{UE}\textsc{v}\text{s}$. 

\paragraph{Noise projection.} To ensure the generated noise can be compatible with event streams, the noise projection has been proposed to process the generated noise. Here, we conduct the ablation study to showcase the necessary of this operation. Note that the polarity of events is normalized into $\{\text{0.0, 0.5, 1.0}\}$, where $\text{0.0/1.0}$ denote the polarity of an event is $\text{-1/+1}$, $\text{0.5}$ means there is no event. As shown in \figref{fig:projection}, there are many noises in the background of the representation frames while removing the projection operation~(original noise). Hence, it's impossible to reconstruct an unlearnable event stream from non-projected representations. We propose \reqref{eq:projection} to project our noise into $\{\text{-0.5, 0, +0.5}\}$.
In \reqref{eq:projection}, $\tau$ is introduced to balance the effectiveness and stealthiness of the injected noise in our $\text{UE}\textsc{v}\text{s}$. A larger $\tau$ can achieve higher stealthiness while corrupting the noise's unlearnable ability. Minimizing the $\tau$ can achieve good effectiveness but the stealthiness would be corrupted.


\section{Conclusion}

This paper proposes the first unlearnable framework for asynchronous event streams~($\text{UE}\textsc{v}\text{s}$), which provides the possibility to protect the event data from being learned by unauthorized models. 
Specifically, a new form of event error-minimizing noise~($\text{E}^2\text{MN}$) is presented to integrate with event data, preventing the model from learning its real semantic features. To ensure the learned noise can be compatible with sparse event data, we propose a projection mechanism to project the $\text{E}^2\text{MN}$ into event friendly counterpart that facility the generation of unlearnable event streams. 
Additionally, a retrieval strategy is proposed to reconstruct the unlearnable event streams from their corresponding unlearnable stacks.
Extensive ablation studies and analysis experiments conducted on four datasets with seven popular DNN models demonstrate the effectiveness, imperceptibility, and robustness of $\text{UE}\textsc{v}\text{s}$. 

\noindent\textbf{Acknowledgement.} Renjie Group is supported by the National Natural Science Foundation of China under Grant No. 62302415, Guangdong Basic and Applied Basic Research Foundation under Grant No. 2022A1515110692, 2024A1515012822, and the Blue Sky Research Fund of HKBU under Grant No. BSRF/21-22/16. Boxin Shi and Peiqi Duan were supported by National Natural Science Foundation of China (Grant No. 62088102, 62402014), Beijing Natural Science Foundation (Grant No. L233024), and Beijing Municipal Science \& Technology Commission, Administrative Commission of Zhongguancun Science Park (Grant No. Z241100003524012).

{
    \small
    \bibliographystyle{ieeenat_fullname}
    \bibliography{main}

\begin{thebibliography}{61}
\providecommand{\natexlab}[1]{#1}
\providecommand{\url}[1]{\texttt{#1}}
\expandafter\ifx\csname urlstyle\endcsname\relax
  \providecommand{\doi}[1]{doi: #1}\else
  \providecommand{\doi}{doi: \begingroup \urlstyle{rm}\Url}\fi

\bibitem[Amir et~al.(2017)Amir, Taba, Berg, Melano, McKinstry, Di~Nolfo, Nayak, Andreopoulos, Garreau, Mendoza, et~al.]{amir2017low}
Arnon Amir, Brian Taba, David Berg, Timothy Melano, Jeffrey McKinstry, Carmelo Di~Nolfo, Tapan Nayak, Alexander Andreopoulos, Guillaume Garreau, Marcela Mendoza, et~al.
\newblock A low power, fully event-based gesture recognition system.
\newblock In \emph{Proc. CVPR}, pages 7243--7252, 2017.

\bibitem[Berlincioni et~al.(2023)Berlincioni, Cultrera, Albisani, Cresti, Leonardo, Picchioni, Becattini, and Del~Bimbo]{berl2023lorenzo}
Lorenzo Berlincioni, Luca Cultrera, Chiara Albisani, Lisa Cresti, Andrea Leonardo, Sara Picchioni, Federico Becattini, and Alberto Del~Bimbo.
\newblock Neuromorphic event-based facial expression recognition.
\newblock In \emph{Proc. CVPR}, 2023.

\bibitem[Cannici and Scaramuzza(2024)]{cannici2024mitigating}
Marco Cannici and Davide Scaramuzza.
\newblock Mitigating motion blur in neural radiance fields with events and frames.
\newblock In \emph{Proc. CVPR}, pages 9286--9296, 2024.

\bibitem[Carlini and Wagner(2017)]{carlini2017towards}
Nicholas Carlini and David Wagner.
\newblock Towards evaluating the robustness of neural networks.
\newblock In \emph{S\&P}, pages 39--57, 2017.

\bibitem[Chen et~al.(2024)Chen, Zhu, Zhang, Hou, Shi, and Wu]{chen2024segment}
Zhiwen Chen, Zhiyu Zhu, Yifan Zhang, Junhui Hou, Guangming Shi, and Jinjian Wu.
\newblock Segment any event streams via weighted adaptation of pivotal tokens.
\newblock In \emph{Proc. CVPR}, pages 3890--3900, 2024.

\bibitem[Dong et~al.(2018)Dong, Liao, Pang, Su, Zhu, Hu, and Li]{dong2018boosting}
Yinpeng Dong, Fangzhou Liao, Tianyu Pang, Hang Su, Jun Zhu, Xiaolin Hu, and Jianguo Li.
\newblock Boosting adversarial attacks with momentum.
\newblock In \emph{Proc. CVPR}, pages 9185--9193, 2018.

\bibitem[Dosovitskiy et~al.(2021)Dosovitskiy, Beyer, Kolesnikov, Weissenborn, Zhai, Unterthiner, Dehghani, Minderer, Heigold, Gelly, Uszkoreit, and Houlsby]{dosovitskiy2020image}
Alexey Dosovitskiy, Lucas Beyer, Alexander Kolesnikov, Dirk Weissenborn, Xiaohua Zhai, Thomas Unterthiner, Mostafa Dehghani, Matthias Minderer, Georg Heigold, Sylvain Gelly, Jakob Uszkoreit, and Neil Houlsby.
\newblock An image is worth 16x16 words: Transformers for image recognition at scale.
\newblock In \emph{ICLR}, 2021.

\bibitem[Duan et~al.(2021{\natexlab{a}})Duan, Wang, Shi, Cossairt, Huang, and Katsaggelos]{duan2021guided}
Peiqi Duan, Zihao~W Wang, Boxin Shi, Oliver Cossairt, Tiejun Huang, and Aggelos~K Katsaggelos.
\newblock Guided event filtering: Synergy between intensity images and neuromorphic events for high performance imaging.
\newblock \emph{TPAMI}, 44\penalty0 (11):\penalty0 8261--8275, 2021{\natexlab{a}}.

\bibitem[Duan et~al.(2021{\natexlab{b}})Duan, Wang, Zhou, Ma, and Shi]{duan2021eventzoom}
Peiqi Duan, Zihao~W Wang, Xinyu Zhou, Yi Ma, and Boxin Shi.
\newblock {EventZoom}: Learning to denoise and super resolve neuromorphic events.
\newblock In \emph{Proc. CVPR}, pages 12824--12833, 2021{\natexlab{b}}.

\bibitem[Duan et~al.(2025)Duan, Li, Yang, Lou, Teng, Zhou, Ma, and Shi]{duan2023eventaid}
Peiqi Duan, Boyu Li, Yixin Yang, Hanyue Lou, Minggui Teng, Xinyu Zhou, Yi Ma, and Boxin Shi.
\newblock Eventaid: Benchmarking event-aided image/video enhancement algorithms with real-captured hybrid dataset.
\newblock \emph{TPAMI}, 2025.

\bibitem[Fei-Fei et~al.(2004)Fei-Fei, Fergus, and Perona]{fei2004learning}
Li Fei-Fei, Rob Fergus, and Pietro Perona.
\newblock Learning generative visual models from few training examples: An incremental bayesian approach tested on 101 object categories.
\newblock In \emph{Proc. CVPRW}, pages 178--178, 2004.

\bibitem[Fu et~al.(2022)Fu, He, Liu, Shen, and Tao]{fu2022robust}
Shaopeng Fu, Fengxiang He, Yang Liu, Li Shen, and Dacheng Tao.
\newblock Robust unlearnable examples: Protecting data privacy against adversarial learning.
\newblock In \emph{ICLR}, 2022.

\bibitem[Gallego et~al.(2020)Gallego, Delbr{\"u}ck, Orchard, Bartolozzi, Taba, Censi, Leutenegger, Davison, Conradt, Daniilidis, et~al.]{gallego2020event}
Guillermo Gallego, Tobi Delbr{\"u}ck, Garrick Orchard, Chiara Bartolozzi, Brian Taba, Andrea Censi, Stefan Leutenegger, Andrew~J Davison, J{\"o}rg Conradt, Kostas Daniilidis, et~al.
\newblock Event-based vision: A survey.
\newblock \emph{TPAMI}, 44\penalty0 (1):\penalty0 154--180, 2020.

\bibitem[Gehrig et~al.(2019)Gehrig, Loquercio, Derpanis, and Scaramuzza]{gehrig2019end}
Daniel Gehrig, Antonio Loquercio, Konstantinos~G Derpanis, and Davide Scaramuzza.
\newblock End-to-end learning of representations for asynchronous event-based data.
\newblock In \emph{Proc. ICCV}, pages 5633--5643, 2019.

\bibitem[Gehrig et~al.(2021)Gehrig, Aarents, Gehrig, and Scaramuzza]{gehrig2021dsec}
Mathias Gehrig, Willem Aarents, Daniel Gehrig, and Davide Scaramuzza.
\newblock Dsec: A stereo event camera dataset for driving scenarios.
\newblock \emph{RAL}, 6\penalty0 (3):\penalty0 4947--4954, 2021.

\bibitem[Goodfellow et~al.(2015)Goodfellow, Shlens, and Szegedy]{goodfellow2014explaining}
Ian~J Goodfellow, Jonathon Shlens, and Christian Szegedy.
\newblock Explaining and harnessing adversarial examples.
\newblock In \emph{ICLR}, 2015.

\bibitem[Gu et~al.(2021)Gu, Sng, Hu, and Yu]{gu2021eventdrop}
Fuqiang Gu, Weicong Sng, Xuke Hu, and Fangwen Yu.
\newblock Eventdrop: Data augmentation for event-based learning.
\newblock In \emph{Proc. IJCAI}, 2021.

\bibitem[Hamann et~al.(2024)Hamann, Ghosh, Martinez, Hart, Kacelnik, and Gallego]{hamann2024low}
Friedhelm Hamann, Suman Ghosh, Ignacio~Juarez Martinez, Tom Hart, Alex Kacelnik, and Guillermo Gallego.
\newblock Low-power continuous remote behavioral localization with event cameras.
\newblock In \emph{Proc. CVPR}, pages 18612--18621, 2024.

\bibitem[He et~al.(2023)He, Zha, and Katabi]{he2023indiscriminate}
Hao He, Kaiwen Zha, and Dina Katabi.
\newblock Indiscriminate poisoning attacks on unsupervised contrastive learning.
\newblock In \emph{ICLR}, 2023.

\bibitem[He et~al.(2016)He, Zhang, Ren, and Sun]{he2016deep}
Kaiming He, Xiangyu Zhang, Shaoqing Ren, and Jian Sun.
\newblock Deep residual learning for image recognition.
\newblock In \emph{Proc. CVPR}, pages 770--778, 2016.

\bibitem[Huang et~al.(2017)Huang, Liu, Van Der~Maaten, and Weinberger]{huang2017densely}
Gao Huang, Zhuang Liu, Laurens Van Der~Maaten, and Kilian~Q Weinberger.
\newblock Densely connected convolutional networks.
\newblock In \emph{Proc. CVPR}, pages 4700--4708, 2017.

\bibitem[Huang et~al.(2021)Huang, Ma, Erfani, Bailey, and Wang]{huang2021unlearnable}
Hanxun Huang, Xingjun Ma, Sarah~Monazam Erfani, James Bailey, and Yisen Wang.
\newblock Unlearnable examples: Making personal data unexploitable.
\newblock In \emph{ICLR}, 2021.

\bibitem[Huang et~al.(2023)Huang, Sun, Zhao, Li, and Su]{huang2023eventpoint}
Ze Huang, Li Sun, Cheng Zhao, Song Li, and Songzhi Su.
\newblock Eventpoint: Self-supervised interest point detection and description for event-based camera.
\newblock In \emph{Proc. WACV}, pages 5396--5405, 2023.

\bibitem[Jiang et~al.(2024{\natexlab{a}})Jiang, Li, Zhang, Deng, and Shi]{jiang2024evhandpose}
Jianping Jiang, Jiahe Li, Baowen Zhang, Xiaoming Deng, and Boxin Shi.
\newblock Evhandpose: Event-based {3D} hand pose estimation with sparse supervision.
\newblock \emph{TPAMI}, 2024{\natexlab{a}}.

\bibitem[Jiang et~al.(2024{\natexlab{b}})Jiang, Zhou, Wang, Deng, Xu, and Shi]{jiang2024complementing}
Jianping Jiang, Xinyu Zhou, Bingxuan Wang, Xiaoming Deng, Chao Xu, and Boxin Shi.
\newblock Complementing event streams and rgb frames for hand mesh reconstruction.
\newblock In \emph{Proc. CVPR}, pages 24944--24954, 2024{\natexlab{b}}.

\bibitem[Jiang et~al.(2023)Jiang, Diao, Wang, Sun, Wang, and Hong]{jiang2023unlearnable}
Wan Jiang, Yunfeng Diao, He Wang, Jianxin Sun, Meng Wang, and Richang Hong.
\newblock Unlearnable examples give a false sense of security: Piercing through unexploitable data with learnable examples.
\newblock In \emph{Pro. ACM MM}, pages 8910--8921, 2023.

\bibitem[Kim et~al.(2021)Kim, Bae, Park, Zhang, and Kim]{kim2021n}
Junho Kim, Jaehyeok Bae, Gangin Park, Dongsu Zhang, and Young~Min Kim.
\newblock N-imagenet: Towards robust, fine-grained object recognition with event cameras.
\newblock In \emph{Proc. ICCV}, pages 2146--2156, 2021.

\bibitem[Kim et~al.(2024)Kim, Cho, and Yoon]{kim2024frequency}
Taewoo Kim, Hoonhee Cho, and Kuk-Jin Yoon.
\newblock Frequency-aware event-based video deblurring for real-world motion blur.
\newblock In \emph{Proc. CVPR}, pages 24966--24976, 2024.

\bibitem[Krizhevsky et~al.(2009)Krizhevsky, Hinton, et~al.]{krizhevsky2009learning}
Alex Krizhevsky, Geoffrey Hinton, et~al.
\newblock Learning multiple layers of features from tiny images.
\newblock 2009.

\bibitem[Lagorce et~al.(2016)Lagorce, Orchard, Galluppi, Shi, and Benosman]{lagorce2016hots}
Xavier Lagorce, Garrick Orchard, Francesco Galluppi, Bertram~E Shi, and Ryad~B Benosman.
\newblock Hots: a hierarchy of event-based time-surfaces for pattern recognition.
\newblock \emph{TPAMI}, 39\penalty0 (7):\penalty0 1346--1359, 2016.

\bibitem[Li et~al.(2017)Li, Liu, Ji, Li, and Shi]{li2017cifar10}
Hongmin Li, Hanchao Liu, Xiangyang Ji, Guoqi Li, and Luping Shi.
\newblock Cifar10-dvs: an event-stream dataset for object classification.
\newblock \emph{Front. Neurosci.}, 11:\penalty0 309, 2017.

\bibitem[Liu et~al.(2023)Liu, Chen, Qu, Zhang, Li, Knoll, and Jiang]{liu2023tma}
Haotian Liu, Guang Chen, Sanqing Qu, Yanping Zhang, Zhijun Li, Alois Knoll, and Changjun Jiang.
\newblock Tma: Temporal motion aggregation for event-based optical flow.
\newblock In \emph{Proc. ICCV}, pages 9685--9694, 2023.

\bibitem[Liu et~al.(2024{\natexlab{a}})Liu, Peng, Zhu, Chang, Zhou, and Yan]{liu2024seeing}
Haoyue Liu, Shihan Peng, Lin Zhu, Yi Chang, Hanyu Zhou, and Luxin Yan.
\newblock Seeing motion at nighttime with an event camera.
\newblock In \emph{Proc. CVPR}, pages 25648--25658, 2024{\natexlab{a}}.

\bibitem[Liu et~al.(2024{\natexlab{b}})Liu, Jia, Xun, Liang, and Cao]{liu2024multimodal}
Xinwei Liu, Xiaojun Jia, Yuan Xun, Siyuan Liang, and Xiaochun Cao.
\newblock Multimodal unlearnable examples: Protecting data against multimodal contrastive learning.
\newblock In \emph{Proc. ACM MM}, pages 8024--8033, 2024{\natexlab{b}}.

\bibitem[Liu et~al.(2024{\natexlab{c}})Liu, Xu, Chen, and Sun]{liu2024stable}
Yixin Liu, Kaidi Xu, Xun Chen, and Lichao Sun.
\newblock Stable unlearnable example: Enhancing the robustness of unlearnable examples via stable error-minimizing noise.
\newblock In \emph{Proc. AAAI}, pages 3783--3791, 2024{\natexlab{c}}.

\bibitem[Liu et~al.(2021)Liu, Lin, Cao, Hu, Wei, Zhang, Lin, and Guo]{liu2021swin}
Ze Liu, Yutong Lin, Yue Cao, Han Hu, Yixuan Wei, Zheng Zhang, Stephen Lin, and Baining Guo.
\newblock Swin transformer: Hierarchical vision transformer using shifted windows.
\newblock In \emph{Proc. ICCV}, pages 10012--10022, 2021.

\bibitem[Ma et~al.(2023)Ma, Paudel, Chhatkuli, and Van~Gool]{ma2023deformable}
Qi Ma, Danda~Pani Paudel, Ajad Chhatkuli, and Luc Van~Gool.
\newblock Deformable neural radiance fields using rgb and event cameras.
\newblock In \emph{Proc. ICCV}, pages 3590--3600, 2023.

\bibitem[Madry et~al.(2018)Madry, Makelov, Schmidt, Tsipras, and Vladu]{madry2018towards}
Aleksander Madry, Aleksandar Makelov, Ludwig Schmidt, Dimitris Tsipras, and Adrian Vladu.
\newblock Towards deep learning models resistant to adversarial attacks.
\newblock In \emph{ICLR}, 2018.

\bibitem[Maqueda et~al.(2018)Maqueda, Loquercio, Gallego, Garc{\'\i}a, and Scaramuzza]{maqueda2018event}
Ana~I Maqueda, Antonio Loquercio, Guillermo Gallego, Narciso Garc{\'\i}a, and Davide Scaramuzza.
\newblock Event-based vision meets deep learning on steering prediction for self-driving cars.
\newblock In \emph{Proc. CVPR}, pages 5419--5427, 2018.

\bibitem[Meng et~al.(2024)Meng, Yi, Yu, Yang, Shen, and Kot]{meng2024semantic}
Ruohan Meng, Chenyu Yi, Yi Yu, Siyuan Yang, Bingquan Shen, and Alex~C Kot.
\newblock Semantic deep hiding for robust unlearnable examples.
\newblock \emph{TIFS}, 2024.

\bibitem[Millerdurai et~al.(2024)Millerdurai, Akada, Wang, Luvizon, Theobalt, and Golyanik]{millerdurai2024eventego3d}
Christen Millerdurai, Hiroyasu Akada, Jian Wang, Diogo Luvizon, Christian Theobalt, and Vladislav Golyanik.
\newblock {EventEgo3D}: {3D} human motion capture from egocentric event streams.
\newblock In \emph{Proc. CVPR}, pages 1186--1195, 2024.

\bibitem[Orchard et~al.(2015)Orchard, Jayawant, Cohen, and Thakor]{orchard2015converting}
Garrick Orchard, Ajinkya Jayawant, Gregory~K Cohen, and Nitish Thakor.
\newblock Converting static image datasets to spiking neuromorphic datasets using saccades.
\newblock \emph{Front. Neurosci.}, 9:\penalty0 437, 2015.

\bibitem[Rebecq et~al.(2017)Rebecq, Horstschaefer, and Scaramuzza]{rebecq2017real}
Henri Rebecq, Timo Horstschaefer, and Davide Scaramuzza.
\newblock Real-time visual-inertial odometry for event cameras using keyframe-based nonlinear optimization.
\newblock In \emph{Proc. BMVC}, pages 16--1, 2017.

\bibitem[Rebecq et~al.(2019{\natexlab{a}})Rebecq, Ranftl, Koltun, and Scaramuzza]{rebecq2019events}
Henri Rebecq, Ren{\'e} Ranftl, Vladlen Koltun, and Davide Scaramuzza.
\newblock Events-to-video: Bringing modern computer vision to event cameras.
\newblock In \emph{Proc. CVPR}, pages 3857--3866, 2019{\natexlab{a}}.

\bibitem[Rebecq et~al.(2019{\natexlab{b}})Rebecq, Ranftl, Koltun, and Scaramuzza]{rebecq2019high}
Henri Rebecq, Ren{\'e} Ranftl, Vladlen Koltun, and Davide Scaramuzza.
\newblock High speed and high dynamic range video with an event camera.
\newblock \emph{TPAMI}, 43\penalty0 (6):\penalty0 1964--1980, 2019{\natexlab{b}}.

\bibitem[Ren et~al.(2023)Ren, Xu, Wan, Ma, Sun, and Tang]{ren2023transferable}
Jie Ren, Han Xu, Yuxuan Wan, Xingjun Ma, Lichao Sun, and Jiliang Tang.
\newblock Transferable unlearnable examples.
\newblock In \emph{ICLR}, 2023.

\bibitem[Rudnev et~al.(2021)Rudnev, Golyanik, Wang, Seidel, Mueller, Elgharib, and Theobalt]{rudnev2021eventhands}
Viktor Rudnev, Vladislav Golyanik, Jiayi Wang, Hans-Peter Seidel, Franziska Mueller, Mohamed Elgharib, and Christian Theobalt.
\newblock Eventhands: Real-time neural {3D} hand pose estimation from an event stream.
\newblock In \emph{Proc. ICCV}, pages 12385--12395, 2021.

\bibitem[Shan et~al.(2020)Shan, Wenger, Zhang, Li, Zheng, and Zhao]{shan2020fawkes}
Shawn Shan, Emily Wenger, Jiayun Zhang, Huiying Li, Haitao Zheng, and Ben~Y Zhao.
\newblock Fawkes: Protecting privacy against unauthorized deep learning models.
\newblock In \emph{USENIX Security}, pages 1589--1604, 2020.

\bibitem[Simonyan and Zisserman(2015)]{simonyan2015very}
K Simonyan and A Zisserman.
\newblock Very deep convolutional networks for large-scale image recognition.
\newblock In \emph{ICLR}, 2015.

\bibitem[Sironi et~al.(2018)Sironi, Brambilla, Bourdis, Lagorce, and Benosman]{sironi2018hats}
Amos Sironi, Manuele Brambilla, Nicolas Bourdis, Xavier Lagorce, and Ryad Benosman.
\newblock {HATS}: Histograms of averaged time surfaces for robust event-based object classification.
\newblock In \emph{Proc. CVPR}, pages 1731--1740, 2018.

\bibitem[Sun et~al.(2024)Sun, Zhang, Zhang, Ma, and Jiang]{sun2024unseg}
Ye Sun, Hao Zhang, Tiehua Zhang, Xingjun Ma, and Yu-Gang Jiang.
\newblock Unseg: One universal unlearnable example generator is enough against all image segmentation.
\newblock In \emph{NeurIPS}, 2024.

\bibitem[Sun et~al.(2022)Sun, Messikommer, Gehrig, and Scaramuzza]{sun2022ess}
Zhaoning Sun, Nico Messikommer, Daniel Gehrig, and Davide Scaramuzza.
\newblock Ess: Learning event-based semantic segmentation from still images.
\newblock In \emph{Proc. ECCV}, pages 341--357, 2022.

\bibitem[Tan and Le(2019)]{tan2019efficientnet}
Mingxing Tan and Quoc Le.
\newblock {EfficientNet}: Rethinking model scaling for convolutional neural networks.
\newblock In \emph{ICML}, 2019.

\bibitem[Vasudevan et~al.(2020)Vasudevan, Negri, Linares-Barranco, and Serrano-Gotarredona]{vasudevan2020introduction}
Ajay Vasudevan, Pablo Negri, Bernabe Linares-Barranco, and Teresa Serrano-Gotarredona.
\newblock Introduction and analysis of an event-based sign language dataset.
\newblock In \emph{FG 2020}, pages 675--682, 2020.

\bibitem[Wang et~al.(2024{\natexlab{a}})Wang, Guo, Li, and Wan]{wang2025event}
Ruofei Wang, Qing Guo, Haoliang Li, and Renjie Wan.
\newblock Event trojan: Asynchronous event-based backdoor attacks.
\newblock In \emph{Proc. ECCV}, pages 315--332, 2024{\natexlab{a}}.

\bibitem[Wang et~al.(2024{\natexlab{b}})Wang, Wang, Tang, Zhu, Jiang, Tian, and Tang]{wang2024event2}
Xiao Wang, Shiao Wang, Chuanming Tang, Lin Zhu, Bo Jiang, Yonghong Tian, and Jin Tang.
\newblock Event stream-based visual object tracking: A high-resolution benchmark dataset and a novel baseline.
\newblock In \emph{Proc. CVPR}, pages 19248--19257, 2024{\natexlab{b}}.

\bibitem[Ye and Wang(2024)]{ye2024ungeneralizable}
Jingwen Ye and Xinchao Wang.
\newblock Ungeneralizable examples.
\newblock In \emph{Proc. CVPR}, pages 11944--11953, 2024.

\bibitem[Zhang et~al.(2024{\natexlab{a}})Zhang, Zhu, and Lam]{zhang2024event}
Pei Zhang, Shuo Zhu, and Edmund~Y Lam.
\newblock Event encryption: Rethinking privacy exposure for neuromorphic imaging.
\newblock \emph{NCE}, 4\penalty0 (1):\penalty0 014002, 2024{\natexlab{a}}.

\bibitem[Zhang et~al.(2025)Zhang, Yin, Wang, Chen, Li, Wang, Lu, and Jia]{zhang2025evsign}
Pengyu Zhang, Hao Yin, Zeren Wang, Wenyue Chen, Shengming Li, Dong Wang, Huchuan Lu, and Xu Jia.
\newblock Evsign: Sign language recognition and translation with streaming events.
\newblock In \emph{Proc. ECCV}, pages 335--351, 2025.

\bibitem[Zhang et~al.(2024{\natexlab{b}})Zhang, Wang, Yang, Shen, and Wen]{zhang2024ev}
Xian Zhang, Yong Wang, Qing Yang, Yiran Shen, and Hongkai Wen.
\newblock Ev-perturb: event-stream perturbation for privacy-preserving classification with dynamic vision sensors.
\newblock \emph{MULTIMED TOOLS APPL}, 83\penalty0 (6):\penalty0 16823--16847, 2024{\natexlab{b}}.

\bibitem[Zubi{\'c} et~al.(2023)Zubi{\'c}, Gehrig, Gehrig, and Scaramuzza]{zubic2023chaos}
Nikola Zubi{\'c}, Daniel Gehrig, Mathias Gehrig, and Davide Scaramuzza.
\newblock From chaos comes order: Ordering event representations for object recognition and detection.
\newblock In \emph{Proc. ICCV}, pages 12846--12856, 2023.

\end{thebibliography}
}

\clearpage
\setcounter{page}{1}
\maketitlesupplementary

\section{Overview}
\label{sec:overview}
\begin{itemize}
    \item \secref{sec:scenario} discusses our \textbf{application scenario} again and shows the \textbf{difference} between image baselines and our unlearnable event streams.
    \item \secref{sec:algorithm} illustrates the detailed explanation of our \textbf{algorithm}.
    \item \secref{sec:class_sample} shows more details of our $\text{E}^2\text{MN}$ and the corresponding \textbf{mixed} counterparts. 

    \item \secref{sec:projection_dis} shows the exploration about the proposed \textbf{projection} strategy.

    \item \secref{sec:baselines} illustrates the details of five kinds of event pollution operations used in our experiments.
    \item \secref{sec:add_exp} add more experiments on \textbf{event representations}, \textbf{time bins}, \textbf{adversarial attack strategies}, \textbf{naive baselines}, \etc.
    
    \item  \secref{sec:dataset} lists the \textbf{dataset} details for N-Caltech101, CIFAR10-DVS, DVS128 Gesture, and N-ImageNet.
    \item \secref{sec:visualization} depicts more \textbf{visualization} results to evaluate the imperceptibility of our $\text{E}^2\text{MN}$.

    \item \secref{sec:impact} discusses the \textbf{social impact} of our $\textsc{UEv}\text{s}$.

    \item \secref{sec:future} lists our \textbf{future work}, including transferability evaluation, generation efficiency, and defense mechanism.

\end{itemize}

\section{Our $\textsc{UEv}\text{s}$}
\label{sec:scenario}

\begin{figure}
    \centering
    \includegraphics[width=\linewidth]{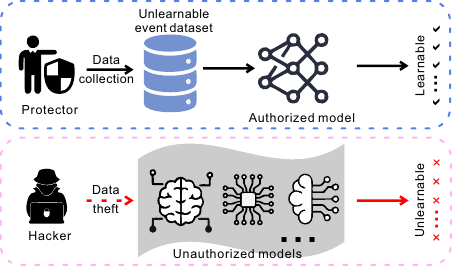}
    \caption{Our scenario of the unlearnable event streams~($\textsc{UEv}\text{s}$). 
    For authorized training, the $\textsc{UEv}\text{s}$ can be effectively used to train downstream models and achieve correct predictions. However, if hackers train their authorized networks without our authority that they cannot achieve the reliable performance.}
    \label{fig:scenario}
\end{figure}

We propose $\text{UE}\textsc{v}\text{s}$ mainly focusing on preventing \textbf{unauthorized} event data usage.
As shown in \figref{fig:scenario}, the protector, \ie, data owner, releases the unlearnable dataset for users, while only the authorized models can learn real semantic features from these data. The hacker's unauthorized models are prevented from learning. 
This mechanism effectively protects the interests of data owners and avoids the privacy leakage caused by data misuse. \figref{fig:scenario} shows the working scenario of our $\text{UE}\textsc{v}\text{s}$.

Unlearnable Examples~(UEs) are proposed to prevent image dataset from unauthorized usage. Compared with UEs, $\text{UE}\textsc{v}\text{s}$ show great challenges. 1) The image perturbation is directly optimized by deep models to ensure its effectiveness and imperceptibility. However, the event data cannot be directly input into deep models to generate the unlearnable version. 
2) Event data shows the binary polarity and asynchronous nature that hinders the noise injection. This sparse property denotes that although we can represent the event stream via image-like features, the generated noise cannot be compatible with event data.
3) Different from image perturbation, event noise should include time stamps that ensure the unlearnable noise is closely aligned with the event data, enhancing imperceptibility.

\begin{figure}
    \centering
    \includegraphics[width=\linewidth]{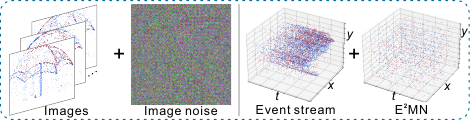}
    \caption{Comparison between the image noise and our $\text{E}^2\text{MN}$~(sample-wise noise).}
    \label{fig:img_event}
\end{figure}

Although $\text{UE}\textsc{v}\text{s}$ and UEs adopt the same core idea: error minimizing loss function, to generate unlearnable noise, using UEs to protect event data is impossible, as shown in \figref{fig:img_event}. The noise is only injected into event representations while the original event data is not protected. Our $\text{UE}\textsc{v}\text{s}$ perturb the original event data that shows better practicality.

\section{$\textsc{UEv}\text{s}$ algorithm}
\label{sec:algorithm}
The algorithm of our $\textsc{UEv}\text{s}$ is shown in Algorithm~\ref{algorithm}. 
%
Here, more detailed explanations about our algorithm are provided.
To generate unlearnable event streams, the protector needs to provide the surrogate model $f'$, target dataset $\mathcal{D}_c$, training step $M$, and classification accuracy $\gamma$. 
$f'$ is employed to calculate the event error-minimizing noise on the dataset $\mathcal{D}_c$. 
$M$ denotes the training iteration of $f'$ in every epoch of noise generation, which is limited since more efforts should be attached on the $\delta$ optimization in $\reqref{eq:emn}$.
Specifically, the training process of $f'$ is listed in {\color{blue}Lines\#3-6}. We randomly sample $M$ batches of event streams from $\mathcal{D}_c$ and incorporate with the noise~(sample-wise noise or class-wise noise) to train the surrogate model. $\mathcal{L}^*$ is our loss function~(\reqref{eq:loss_surrogate}), consisting of the cosine similarity loss function and cross-entropy loss function. The final loss is calculated as:

\begin{equation}
\resizebox{0.9\linewidth}{!}{$
    \begin{split}
    &loss=\lambda_1([1+ { \frac{f'(\mathcal{R}(\mathcal{E}))_{conv} \bm{\cdot} f'(\mathcal{R}(\mathcal{E})+\delta)_{conv} }{||f'(\mathcal{R}(\mathcal{E}))_{conv}|| \times ||f'(\mathcal{R}(\mathcal{E})+\delta)_{conv}||}})]/2) + \\
    &\lambda_2 (-[l_i log f'(\mathcal{R}(\mathcal{E})+\delta) + (1-l_i)log(1-(f'(\mathcal{R}(\mathcal{E})+\delta))]),
    \label{eq:training}
    \end{split}$}
\end{equation}
where $(\cdot)_{conv}$ denotes the convolution features extracted by the last convolution layer of surrogate model $f'$, $B$ indicates the batch size, $\mathcal{R}$ means event representation, converting an event stream into the event stack. \reqref{eq:training} illustrates the training pipeline of $f'$ on a batch of event streams.
It is not required to calculate the cost of $f'(\mathcal{R}(\mathcal{E}))$ because this would cause the surrogate model to focus on learning the real semantic features, thereby preventing the optimization of our unlearnable noise.

After training, we generate the event error-minimizing noise for entire $\mathcal{D}_c$ according to \reqref{eq:adversarial-noise}, as shown in {\color{blue}Lines\#7-10}. $Clip(\cdot)$ denotes clipping those noise that exceed $-\epsilon$ or $+\epsilon$ back to this region. The noise generation and surrogate model training would be terminated once the classification accuracy tested under $\delta$ is higher than $\gamma$. This termination demonstrates that the generated noise can effectively guide the model to conduct predictions without relying on image semantics. Therefore, the noise $\delta$ is able to prevent the unauthorized model from learning informative knowledge from our data.

Due to the special characteristic of event data, our noise $\delta$ is generated based on the event stack. It's necessary to conduct event reconstruction to generate the unlearnable event stream from its corresponding unlearnable event stack~({\color{blue}Lines\#13-17}).
To ensure the $\delta$ can be compatible with event data, we propose a projection strategy~$\mathbf{P}(\cdot)$ to sparsify the noise into $\{-0.5, 0, +0.5\}$\footnote{In image area, the generated noise is directly added on the images, rendering the unlearnable examples via modifying the pixel values. However, for event data, which consists solely of binary events, we can only transform the event stream into an unlearnable one on the event level via deleting events or inserting new events. Therefore, we define the projected values as $\{-0.5, 0, +0.5\}$ to be compatible with event stacks~($\{0, 0.5, 1.0\}$).}. 
Then, we integrate the projected noise with an event stack and clip it into $[0,1]$ to generate the unlearnable event stack. Detailed noise embedding process is shown in \tabref{tab:confusion}. Finally, a retrieval strategy $\mathcal{R}'$ is proposed to search the compressed time stamps from the original event streams to achieve the reconstruction. Based on this algorithm, our valuable event data can be protected well that prevents unauthorized data exploitation, as shown in \figref{fig:scenario}.

\begin{figure}[t]
    \centering
    \includegraphics[width=\linewidth]{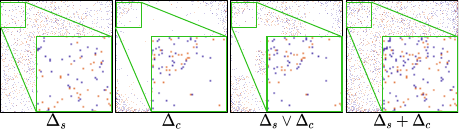}
    \caption{Illustrative examples of our $\text{E}^2\text{MN}$.}
    \label{fig:noise}
\end{figure}

\section{Class-wise and sample-wise noise}
\label{sec:class_sample}
Our $\text{E}^2\text{MN}$ consists of two kinds of noise: class-wise noise and sample-wise noise, which are all generated based on \reqref{eq:emn}. The sample-wise noise is generated case by case, which leads to every generated noise being only workable for the single event stream. This limits the practicality of sample-wise noise, especially in the dataset scale grows or new event streams are captured. Hence, an alternative way is class-wise noise, which is generated class by class. It means that a kind of noise can be injected into different event streams sampled from the sample class. Another advantage is that class-wise noise consumes less memory than sample-wise noise in noise optimization. Although two kinds of noises achieve similar performance in \tabref{tab:experiment}, class-wise noise achieves better stealthiness than sample-wise noise as shown in \figref{fig:result} and \tabref{tab:visual}. 

Considering their respective advantages, we combine the two noises to explore whether it can bring more benefits.
We have proposed union and addition operations in \secref{sec:ablation} to evaluate. 
For $\Delta_s\vee\Delta_c$, we randomly choose $\Delta_s$ or $\Delta_c$ to protect the event. The form of this kind of noise resembles both of them because only a simple random sampling operation is employed.
For $\Delta_s+\Delta_c$, we fuse two kinds of noise by element-wise addition to perturb the event stream. 
As shown in \figref{fig:noise}, this configuration increases the number of perturbations introduced into the event data, resulting in higher effectiveness~(see E4 of \tabref{tab:abl_exp}).

\begin{table}[]
    \centering
    \caption{Confusion matrix of embedding the noise~($\text{E}^2\text{MN}$) into an event stack~(E. stack). The final unlearnable event stack would be clipped into $[0,1]$.}
    \resizebox{\linewidth}{!}{
    \begin{tabular}{ccccc}
    \toprule
    &&\multicolumn{3}{c}{$\delta$} \\
    \cline{3-5}
    &&$-0.5$&$0$&$+0.5$ \\
    \midrule
    \multirow{3}{*}{\rotatebox{90}{E. stack}} & $0$ &original event & original event & event deletion\\
    & $0.5$ &event generation& no event & event generation\\
    & $1.0$ &event deletion& original event & original event\\
    \bottomrule
    \end{tabular}}
    
    \label{tab:confusion}
\end{table}

\begin{figure}[h]
    \centering
    \includegraphics[width=\linewidth]{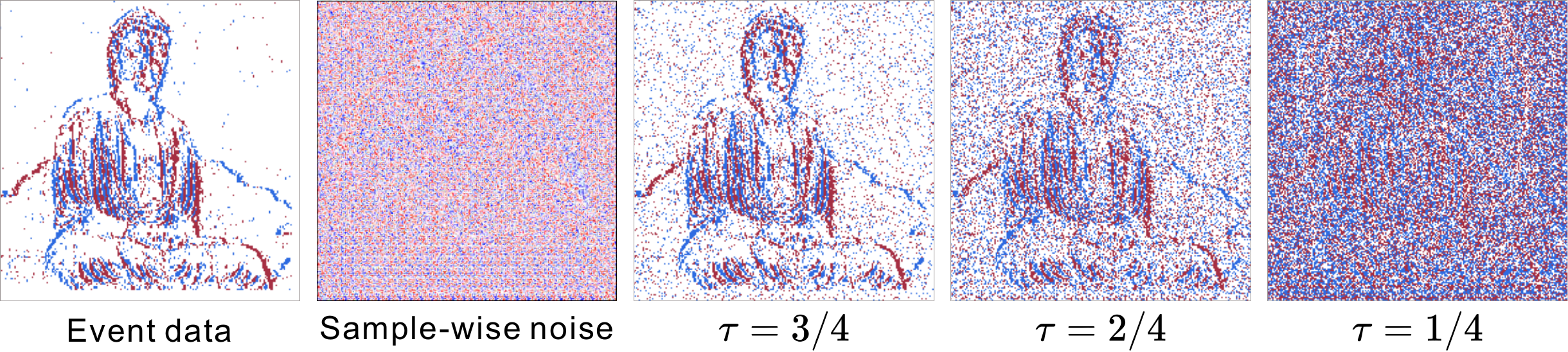}
    \caption{Visualization of event data, sample-wise noise, and the corresponding unlearnable event streams with different projection parameters $\tau$.}
    \label{fig:projection_pi}
\end{figure}

\section{Projection discussion}
\label{sec:projection_dis}

Our projection strategy is designed to sparsify the noise $\delta$ to ensure compatibility with event stacks.
We illustrate a detailed confusion matrix of the noise embedding in \tabref{tab:confusion}. 
If $\delta=-0.5$ is added to a negative event (0), the event keeps its original value. If added to a positive event (1), the event is deleted. A new event with negative polarity is created when adding $-0.5$ to the pixel value 0.5 sampled from the event stack.
In this section, we showcase the importance of the parameter $\tau$ in \reqref{eq:projection}. As discussed in \cite{huang2021unlearnable}, the added noise should be imperceptible to human eyes and does not affect the normal data utility. Hence, we introduce the parameter $\tau$ to balance the imperceptibility and unlearnability of our $\text{E}^2\text{MN}$. As shown in \figref{fig:projection_pi}, the larger $\tau$ can lead to better imperceptibility but the less unlearnable noise has been introduced, which harms the unlearnability. We have tested our sample-wise noise with $\tau=7/8$ on the N-Caltech101 dataset and obtained the accuracy of ResNet18 by 4.88, which is higher than the accuracy~(0.52) tested by $\tau=3/4$. This demonstrates the great challenges in balancing the imperceptibility and unlearnability of our $\text{E}^2\text{MN}$.

\begin{figure*}[h]
    \centering
    \includegraphics[width=\linewidth]{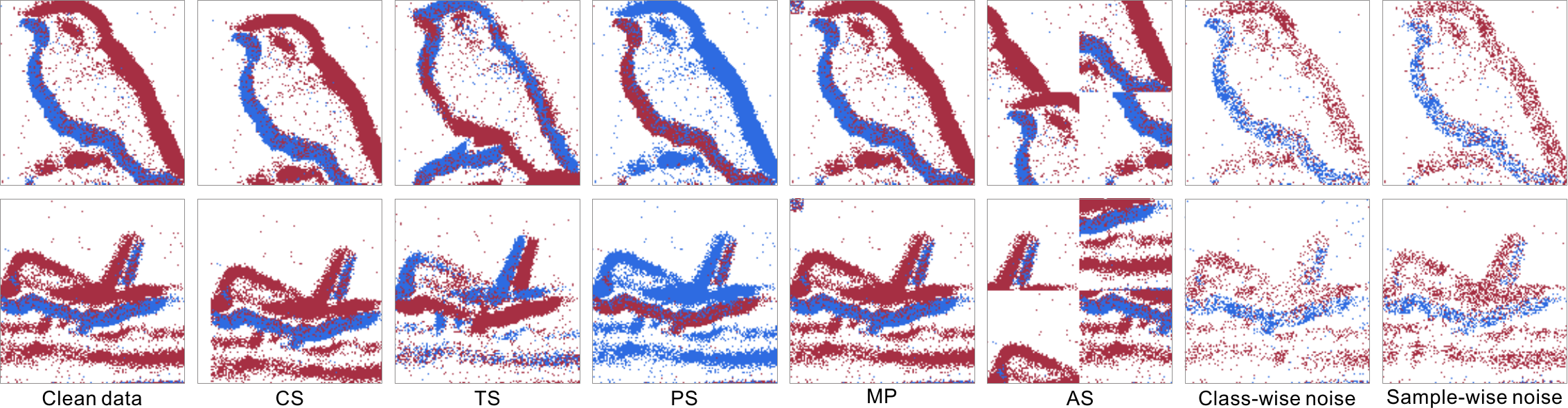}
        \caption{Visualization results of various noise forms on the CIFAR10-DVS dataset~\cite{li2017cifar10}. Blue/Red points denote the events with $p=+1$/$-1$. 
    Our noise $\text{E}^2\text{MN}$ does not introduce much noise in the background region, which maintains good imperceptibility.
    }
    \label{fig:cifar}
\end{figure*}

\begin{figure*}[h]
    \centering
    \includegraphics[width=\linewidth]{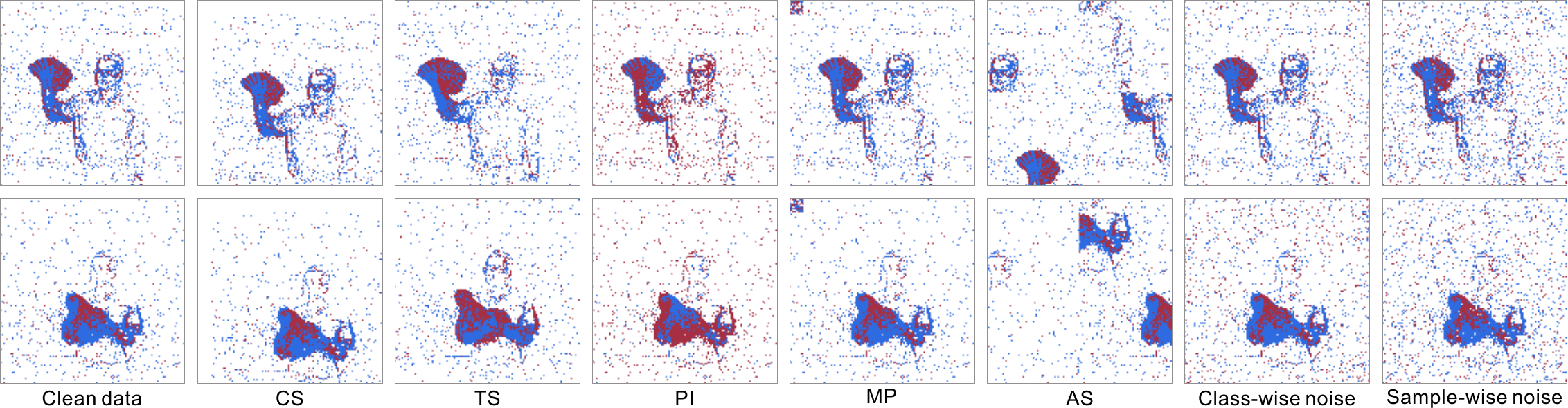}
        \caption{Visualization results of various noise forms on the DVS128 Gesture dataset~\cite{amir2017low}. Blue/Red points denote the events with $p=+1$/$-1$. 
    Our noise $\text{E}^2\text{MN}$ does not corrupt the target objects, and the noise distributed in the background region appears quite realistic.
    }
    \label{fig:gesture}
\end{figure*}

\section{Baseline setting}
\label{sec:baselines}
To further evaluate the effectiveness of our $\textsc{UEv}\text{s}$, we introduce the straightforward event distortions as our baselines, which are inspired from event data augmentation~\cite{gu2021eventdrop} and backdoor attacks~\cite{wang2025event}. Data augmentation is proposed to enrich the training data for improving the model's performance, which usually employs data distortion operations to augment the sample. Generally, the quality of these augmented samples is lower than the original ones. Therefore, we propose simple coordinate shifting~(CS), timestamp shifting~(CS), polarity inversion~(PI), and area shuffling~(AS) based on \cite{gu2021eventdrop} to corrupt the event streams for preventing unauthorized usage.
Additionally, we also propose the manual pattern~(MP) based on backdoor attacks~\cite{wang2025event} to perturb our event streams. 
We inject a pre-defined pattern for those event streams sampled from the same class to prevent unauthorized data usage, which can be viewed as a class-wise noise.
According to \tabref{tab:experiment}, 
we can find that compromising the quality of our event datasets can degrade the performance of downstream models. However, the unlearnability is rather limited and does not prevent the downstream models from learning informative knowledge.

\section{Additional experiments}
\label{sec:add_exp}
\paragraph{Various event representations.} 
In our main experiments, we adopt the voxel-grid event stack as our event representation to evaluate the effectiveness. To test the generalizability of our $\text{UE}\textsc{v}\text{s}$ among different representations, the event frame~(\textbf{EF})~\cite{rebecq2017real} and Time surface~(\textbf{TS})~\cite{sironi2018hats} are adopted. 
As shown in the \textbf{E1} of \tabref{tab:exp}, our $\textsc{UEv}$s can still prevent unauthorized event data usage under \textbf{EF} and \textbf{TS} representations, showing high robustness and generalizability.

\paragraph{Generalizability.} According to \cite{duan2021eventzoom}, we set the time bin to 16 to represent the event stream. To evaluate the generalizability of our $\textsc{UEv}$s on different time intervals, we change the size of $\Delta_t$ to $\mathbf{0.5\times}$ and $\mathbf{2\times}$ to conduct ablation studies. as shown in \textbf{E2} of \tabref{tab:exp}, our $\textsc{UEv}$s still shows high protection ability to prevent the unauthorized event data usage.

\paragraph{Diverse Adversarial attacks.} Apart from the PGD~\cite{madry2018towards} and FGSM~\cite{goodfellow2014explaining} attacks, we add new adversarial attack methods: C\&W~\cite{carlini2017towards} and MIFGSM~(MIF.)~\cite{dong2018boosting}. 
CW attack is an optimization-based method that crafts minimal perturbations to mislead neural networks while remaining imperceptible.
MIFGSM is an iterative adversarial attack that enhances the basic FGSM by incorporating the momentum. Compared with MIFGSM, CW attack achieves better unlearnable functionality.
As shown in \textbf{E3} of \tabref{tab:exp}, our method can still achieve reliable unlearnable performance while adopting different adversarial attacks.

\begin{table}[h]
    \centering
    \caption{Quantitative results tested by Res50 on N-Caltech101.}
    \resizebox{0.9\linewidth}{!}{
    \begin{tabular}{ccc|cc|cc|c}
    \toprule
    &\multicolumn{2}{c}{E1}&\multicolumn{2}{c}{E2}&\multicolumn{2}{c}{E3}&E4 \\

    &EF&TS&0.5$\Delta_t$&$2\Delta_t$&C\&W&MIF.&UEs\\
    \midrule
    $\Delta_c$ &5.17 &10.09 &5.46 &4.94 &8.73 &15.43 &5.51 \\
    $\Delta_s$ & 8.85&16.48&5.17&5.05&12.15 &14.47 &18.38 \\
     \bottomrule
    \end{tabular}}
    \label{tab:exp}
    \vspace{-10pt}
\end{table}

\paragraph{Naive image baselines.}
Image-based methods are unable to directly secure event data due to data differences, which can only safeguard the corresponding event representation.
In \textbf{E4} of \tabref{tab:exp}, although the image approach, UEs~\cite{huang2021unlearnable}, performs well in event representations, it cannot prevent malicious users from misusing the \textbf{original event}.

\begin{figure}[h]
    \centering
    \vspace{-5pt}
    \includegraphics[width=\linewidth]{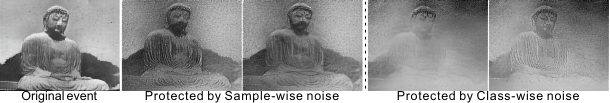}
 
    \caption{Visualization frames reconstructed by E2VID \cite{rebecq2019events}.}
    \vspace{-5pt}
    \label{fig:buddha}
   
\end{figure}

\paragraph{Event to image reconstruction.}
We employ an event-to-image method~(E2VID~\cite{rebecq2019events}) to evaluate the generazibility. As shown in \figref{fig:buddha}, our method prevents E2VID~\cite{rebecq2019events} from reconstructing details from the protected event data, thereby providing solutions for privacy preserving.

\paragraph{Unlearnable cluster.}
We extend our class-wise noise into a cluster-wise one. We \ding{182} employ K-Means+ResNet50 to cluster the N-Caltech101 into \textbf{10} classes; \ding{183} train a surrogate model on 10 classes to generate \textbf{cluster-wise noise}; \ding{184} train ResNet18 on the whole classes (\textbf{101}) with cluster-wise noise. The cluster version of our method can reduce the classification accuracy from $0.787$ to $0.189$.

\section{Dataset details}
\label{sec:dataset}
To evaluate the effectiveness of our method, we employ four popular event-based datasets in our experiments, including N-Caltech101~\cite{orchard2015converting}, CIFAR10-DVS~\cite{li2017cifar10}, DVS128 Gesture~\cite{amir2017low}, and N-ImageNet~\cite{kim2021n}.
N-Caltech101 is the neuromorphic version of the image dataset, Caltech101~\cite{fei2004learning}, which has $101$ classes and $4356$, $2612$, and $1741$ samples for training, validation, and testing, respectively. 
CIFAR10-DVS is generated based on image datasets CIFAR-10~\cite{krizhevsky2009learning}, where the training set, validation set and testing set contain $7000$, $1000$, and $2000$ samples, respectively.
DVS128 Gesture contains $11$ classes from $29$ subjects under $3$ illumination conditions, which has $1176$ training samples and $288$ testing samples.
The N-ImageNet (mini) dataset is derived from the ImageNet dataset. It utilizes an event camera to capture RGB images shown on a monitor. This dataset includes 100 object classes, with each class having 1,300 streams for training and 50 streams for validation. Details of each dataset are shown in \tabref{tab:dataset}.

\begin{table}[h]
    \centering
    \caption{Details of four used datasets in our experiments.}
    \resizebox{\linewidth}{!}{
    \begin{tabular}{lcccc}
    \toprule
    & N-Caltech101 & CIFAR10-DVS & DVS128 Gesture & N-ImageNet \\
    \midrule
    Type & Simulated & Simulated & Real & Simulated\\
    Calsses& 101 & 10 & 11 & 1000\\
    Resolution&$180\times 240$&$128\times 128$ &$128\times 128$&$480\times640$\\
    Event Camera&ATIS camera&DVS128 &DVS128&Samsung DVS Gen3\\
    Train&4536&7000 &1176&130000\\
    Val&2612&1000 &288&50000\\
    Test &1741&2000 &288&50000\\
    
    \bottomrule
    \end{tabular}}

    \label{tab:dataset}
\end{table}

\section{Visualization comparison}
\label{sec:visualization}
To evaluate the imperceptibility of our $\text{E}^2\text{MN}$, we show more visualization results in \figref{fig:cifar} and \figref{fig:gesture}. In \figref{fig:cifar}, we sample event streams from the CIFAR10-DVS dataset~\cite{li2017cifar10} to generate the unlearnable ones via five straightforward distortions and our two kinds of noise. 
It’s clear that there is an $x$-$y$ offset in the unlearnable event streams generated by CS compared to the clean data. TS alters time stamps to pollute the input event streams, resulting in a noticeable difference between the distorted samples and the clean data. PI reverses the polarity of the event stream to degrade the quality, thereby reducing the quality of the training data. AS rearranges the event data at the block level to produce unlearnable data, which exhibits low imperceptibility.
Our class-wise noise and sample-wise noise perturb the event stream with imperceptible noise that shows better invisibility than other comparison methods. The visualization results sampled from DVS128 Gesture dataset~\cite{amir2017low} are shown in \figref{fig:gesture}. It's clear that our $\text{E}^2\text{MN}$ achieves the best unlearnability while maintaining good imperceptibility.

\begin{figure}[h]
    \centering
    \includegraphics[width=\linewidth]{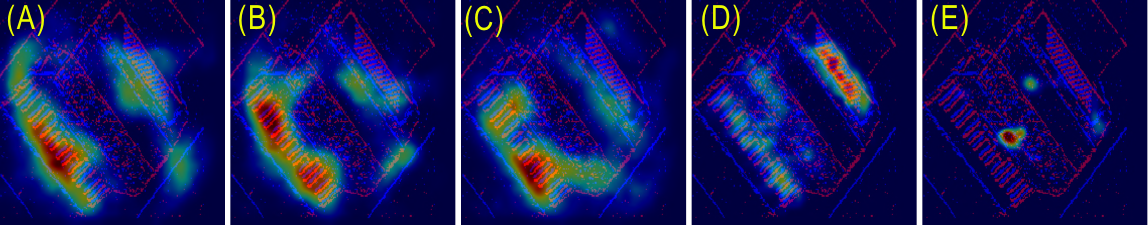}
    \caption{GradCAM of event distortions and our methods.}
    \label{fig:cam}
    \vspace{-5pt}
\end{figure}

To visualize the influence caused by our $\text{E}^2\text{MN}$, we employ GradCAM to highlight several unlearnable event streams in \figref{fig:cam}. 
Figures (A), (B), (C), (D), and (E) are rendered by A-shuffle, M-pattern, P-inverse, Class-wise noise, and sample-wise noise, respectively. Our method builds a shortcut between the input samples and labels that suppresses the model learning semantic features, resulting in a lower response on the foreground regions.

\section{Social impact}
\label{sec:impact}
The social impact of $\textsc{UEv}\text{s}$ is multifaceted, addressing key issues around event data security, privacy, and ethics.
By making event datasets unlearnable, our method helps protect individuals' data from being used without authorization. This is particularly important in an era where event data privacy concerns are paramount, and unauthorized data usage can lead to significant privacy breaches. 
The method provides a robust protection way for event data owners, ensuring that their event streams cannot be exploited by unauthorized entities. This fosters greater trust in event data sharing. 
With the application of $\textsc{UEv}\text{s}$, there is a push towards more ethical event data practices. Entities will need to obtain proper authorization and consent before using event data, promoting a culture of respect for data ownership and user rights.

Overall, our $\textsc{UEv}\text{s}$ contributes significantly to the advancement of secure and trustworthy data sharing, promoting a safer and more ethical event data ecosystem.

\section{Future work}
\label{sec:future}
$\textsc{UEv}\text{s}$ is the first method designed for generating unlearnable event streams, which provides a possible solution to prevent the unauthorized usage of our valuable event data. 
We mainly focus on studying the unlearnability of event streams in the main paper, while causing some limitations in terms of transferability evaluation, generation efficiency, and defense mechanism. 
To address these issues beyond our research topic in this work, we will explore the following directions in the future:
\begin{itemize}
    \item Transferability evaluation: We plan to extend the evaluation of our method from classification task to other event datasets and event vision tasks, enhancing the transferability. This comprehensive testing is helpful to demonstrate the effectiveness of $\textsc{UEv}\text{s}$ that promote the trustworthy event data sharing. However, the effectiveness of $\textsc{UEv}\text{s}$ may be decreased across different vision tasks. A possible solution is that adopt the foundation model as our surrogate model to calculate the event error-minimizing noise, which is trained with a large amount of data that has strong generalization capabilities. It can be applied to a variety of tasks without training independent models for each specific task.

    \item Generation efficiency: According to our experiments, we find that the efficiency of the sample-wise noise generation depends on the scale of the event dataset. The larger the scale of the event dataset, the lower the efficiency of the sample-wise noise generation. We propose addressing this issue via a noise generator. We train this generator with the surrogate model jointly, which aims to enable the generated noise to minimize the cost of the surrogate model. This training pipeline avoids storing the generated medium noise, which saves efficiency significantly.
    Once the optimization has been finished, we can employ this generator to generate sample-wise noise for each input sample with high efficiency. 

    \item Defense mechanism: 
    It's crucial to investigate potential defense mechanisms against the generation of malicious unlearnable event streams. If we release our unlearnable dataset online, hackers could manipulate these samples to force their models to learn the information. By understanding possible defense mechanisms, we can develop more reliable methods for creating unlearnable event datasets, thereby preventing unauthorized usage. Furthermore, elucidating these defense mechanisms can help users improve the dataset quality, ultimately saving training time and computing resources.
\end{itemize}

\end{document}